\documentclass[twocolumn]{openjournal}

\usepackage{graphicx}   
\usepackage{amsmath}    
\usepackage{amssymb}    
\usepackage{hyperref}   
\usepackage{lipsum}     
\usepackage{color}
\usepackage[normalem]{ulem}

\newcommand{\MyRed}{\color [rgb]{0.9,0,0.1}}

\def\vpc#1{{\bf\MyRed[VP: #1]}}

\begin{document}

\title{decaying dark matter halo abundance from a revised spherical collapse model}\def\vpc#1{{\bf\MyRed[VP: #1]}}
\author{Thomas Montandon}
\author{Vivian Poulin}

\affiliation{Laboratoire Univers et Particules de Montpellier, Universit\'e de Montpellier/CNRS, place E. Bataillon, cc072, 34095 Montpellier, France}

\author{Oliver Hahn}
\affiliation{Department of Astrophysics, T\"urkenschanzstraße 17, 1180 Vienna, Austria\\Department of Mathematics, Oskar-Morgenstern-Platz 1, 1090 Vienna, Austria}
\author{Jozef Bucko}
\affiliation{Department of Physics, ETH Zurich, Wolfgang-Pauli-Strasse 27, CH-8093 Zurich, Switzerland}

\author{Aurel Schneider}
\affiliation{Department of Astrophysics, University of Zurich, Winterthurerstrasse 190, CH-8057 Zurich, Switzerland}

\altaffiltext{1}{Corresponding author: \href{mailto:email@example.com}{thomas.montandon@umontpellier.fr}}

\begin{abstract}
We present a semi-analytical framework for the halo mass function (HMF) in decaying dark matter (DDM) cosmologies, in which dark matter decays into a massive daughter particle inheriting a velocity kick $v_k$ and a massless dark radiation component. Building on the Press--Schechter formalism, we encode the DDM physics through a spherical collapse model that explicitly tracks the decay-induced mass loss, yielding a modified, mass-dependent critical collapse threshold $\delta_c(M_0)$ and a mapping $M_{\rm coll}(M_0)$ between the initial Lagrangian mass and the collapsed halo mass. The critical threshold exhibits a characteristic transition between two analytically tractable plateaus: a large-mass limit, where all daughter particles are retained by the halo, and a small-mass limit, where all daughters escape and the collapse is equivalent to that of a dark matter species decaying entirely into dark radiation, making $\delta_c$ independent of $M_0$ and $v_k$. We provide semi-analytical results and fits for both limits and a fitting formula for the transition, whose single free parameter $M_1 \propto v_k^3\,\tilde\Gamma^{-1/2} t_{\rm ta}$ has a transparent physical interpretation: it is the mass scale at which the kick velocity equals the halo orbital velocity. We validate our predictions against a suite of N-body simulations at $z=0$ and $z\approx 1$, finding good agreement across models spanning mild to strong HMF suppression relative to $\Lambda$CDM. Residual deviations for the largest kick velocities at $z=0$ are observed. Via a halo-by-halo comparison between simulations, we trace the discrepancy to the definition of the halo mass when daughter orbits extend beyond the halo boundary. The resulting fitting functions for $\delta_c(M_0,\Gamma,v_k)$ and $M_{\rm coll}(M_0)$ provide an efficient and accurate route to DDM constraints from current and forthcoming probes of the halo mass function.
\end{abstract}

\section{Introduction}
Over the past decades, the $\Lambda$CDM cosmological model has emerged as the standard paradigm for describing the Universe, providing an excellent fit to a wide range of observations spanning both early and late cosmic times. However, despite its empirical success, the fundamental nature of cold dark matter (CDM) remains unknown, and tensions have emerged as the precision of cosmological data has improved. On cosmological scales, a known discrepancy is the so-called $S_8$ tension, referring to the disagreement between the amplitude of matter fluctuations inferred from high-redshift CMB data (e.g. $S_8 = 0.825 \pm 0.011$ from \textit{Planck} \citep{Aghanim:2018eyx}) and direct measurements at low redshift from weak lensing surveys like \texttt{KiDS} \citep{Kuijken:2019gsa,Giblin:2020quj, Busch:2022pcx}, \texttt{DES} \citep{DES:2021bvc}, and \texttt{HSC} \citep{Hamana:2019etx}. While recent analyses like \texttt{DES-Y3} \citep{DES:2021wwk} and \texttt{KiDS-Legacy} \citep{Stolzner:2025htz} mitigate this tension, the latest $S_8$ measurements from \texttt{DES-Y6} still show a significant discrepancy with {\it Planck} \citep{DES:2026fyc}. Upcoming data from the Vera Rubin Observatory \citep{LSST:2008ijt}, \textit{Euclid} \citep{Euclid:2024yrr}, the Nancy Grace Roman Space Telescope \citep{Spergel:2015sza}, and the Chinese Space Station Telescope \citep{Gong:2019yxt} will provide definitive answers regarding small-scale physics.

Various physical explanations have been proposed in the literature, assuming this discrepancy originates from real physics rather than systematics. Baryonic feedback combined with nonlinear structure formation \citep{Tan:2022wob, Amon:2022azi, Arico:2023ocu} are among the most discussed. Perhaps more exciting, however, is the possibility that the tension points to entirely new physics in the dark sector, motivating alternatives to standard CDM \citep{Schneider:2019xpf, Heimersheim:2020aoc, Joseph:2022jsf, Poulin:2022sgp, Ferlito:2022mok} or Primordial non-Gaussianities \citep{Stahl:2024stz}. Among these, decaying dark matter (DDM) scenarios \citep{Enqvist:2015ara, Enqvist:2019tsa, Murgia:2017lwo, Abellan:2021bpx, DES:2020mpv, Choi:2021uhy, Tanimura:2023bkh, Bucko:2022kss} make the simple and natural assumption that dark matter may be unstable on cosmological timescales, an assumption/prediction shared by many particle physics-motivated models \citep{Hambye:2010zb, Abazajian:2012ys, Drewes:2016upu, Doroshkevich:1984gw, Doroshkevich:1989bf, Khlopov:1995pa, Berezinsky:1991sp, Covi:1999ty, Kim:2001sh, Chou:2003wx, Feng:2003uy, Ghosh:2020ipv, Dutta:2022wuc, Fuss:2024dam}.

A particularly interesting variant is the DDM model with one massive daughter\footnote{Sometimes referred to as ``two-body decays'', meaning DM decays into particles of two different masses, while the extension ``one-body decays'' refer to decay into two massless particles.}, in which a fraction $f_{\rm DDM}$ of the CDM decays into a massive daughter particle and a massless component (dark radiation). In this work, we only consider  $f_{\rm DDM} = 1$. By energy-momentum conservation, the massive daughter inherits a velocity kick $v_k$, which drives a suppression of the matter power spectrum on small scales at late times. \citet{Wang:2012eka} and \citet{Abellan:2021bpx} investigated the phenomenology of this model using linear perturbation theory, while \citet{Cheng:2015dga} and \citet{Bucko:2023eix} extended the analysis into the nonlinear regime with N-body simulations. Constraints on the DDM parameters have been derived from CMB data and weak lensing \citep{Nygaard:2020sow, Holm:2022kkd, Simon:2022ftd, Bucko:2023eix, Montandon:2025xpd}, as well as from Lyman-$\alpha$ forest observations \citep{Wang:2013rha, Fuss:2022zyt}.

Several works have studied the impact of DDM on small-scale structure using N-body simulations \citep{Peter:2010sz, Wang:2014ina, Cheng:2015dga}, showing that velocity kicks heat and disrupt low-mass halos, flatten inner density profiles, and suppress the subhalo mass function. The latest constraints from subhalo counts come from \citet{DES:2022doi}, who used Milky Way satellite counts to constrain the DDM lifetime and velocity kick.

Regarding the halo mass function, cluster surveys have delivered increasingly precise measurements across a wide mass range, enabling tight cosmological constraints \citep{Zubeldia:2020knz, DES:2020cbm, Salvati:2021gkt, Sunayama:2023hfm, SPT:2024qbr, Aymerich:2024fim}, with the eROSITA all-sky survey \citep{Artis:2024zag} representing the current state of the art and opening new avenues to constrain DDM from cluster number counts. On the theoretical side, \citet{Cheng:2015dga} provided a first simulation-based fit to the DDM HMF, though limited in parameter space $v_k \leq 200~\mathrm{km/s}$ for two lifetimes, while \citet{Nadler:2025yni} recently developed a semianalytic framework for the subhalo mass function and halo profiles in DDM cosmologies. 

In this work, we extend the Press--Schechter (PS) formalism \citep{Press:1973iz, Bond:1990iw} to incorporate the DDM physics developed in \citet{Nadler:2025yni}. This framework provides a statistical connection between the abundance of dark matter haloes and the underlying matter density field, with the critical overdensity $\delta_c$ playing a central role as the collapse barrier. To evaluate $\delta_c$, we use the spherical collapse model \citep{1980lssu.book.....P, Gunn:1972sv}, which in $\Lambda$CDM yields the well-known nearly universal result $\delta_c \approx 1.686$ \citep{Lahav:1991wc, Eke:1996ds, Cooray:2002dia}. The PS prediction with this threshold, however, overpredicts low-mass halo abundances and underpredicts massive ones compared to simulations \citep{Sheth:1999mn}. This was improved by \citet{Sheth:1999mn, Sheth:2001dp} by moving from spherical to ellipsoidal collapse. Deeper theoretical grounding was later provided by the Excursion Set Peaks formalism \citep{Maggiore:2009rv, Maggiore:2009rw, Musso:2012qk, Paranjape:2012ks}, which reproduces the ellipsoidal suppression without ad-hoc parameter tuning. Further work has challenged density as the fundamental collapse variable \citep{Musso:2019zmr}, and highlighted significant stochasticity in the Lagrangian mass-radius relation \citep{Wislocka:2025kvd}.

On the simulation side, precision fits to the HMF have been provided by a series of increasingly accurate N-body calibrations \citep{Jenkins:2000bv, VIRGO:2001szp, Reed:2003sq, Crocce:2009mg, Pillepich:2008ka, Tinker:2008ff, Angulo:2012ep, Watson:2012mt, Bocquet:2015pva, Despali:2015yla}, with \citet{Tinker:2008ff} remaining among the most widely used, and more recently supplemented by machine-learning-based emulators \citep{Bocquet:2020tes, Buisman:2025qaz}. None of these frameworks, however, account for the modified collapse dynamics inherent to DDM cosmologies. This is the goal of the present work: building on \citet{Nadler:2025yni}, we extend their approach to the halo mass function through a spherical collapse model that self-consistently incorporates DDM physics, and provide an analytical fitting framework for the mass-dependent critical density $\delta_c(M)$ that can be used directly in cosmological analyses. Note that in practice, we use the standard Sheth–Tormen multiplicity function, which captures the effects of ellipsoidal collapse.

The paper is organised as follows. In Sec.~\ref{sec:theory}, we present the theoretical framework: the modified Press--Schechter formalism and the DDM spherical collapse model from which the mass-dependent critical density $\delta_c(M)$ is derived. In Sec.~\ref{sec:Critical_density}, we systematically study $\delta_c(M)$, derive analytical expressions for its large- and small-mass limits, and provide semi-analytical fitting formula that can be used directly in place of solving the full system of equations. In Sec.~\ref{sec:simulations}, we describe the N-body simulations and halo-finding procedure used throughout this work. In Sec.~\ref{sec:validation}, we validate our predictions against the N-body simulations and discuss the role of the halo mass definition in the comparison. We conclude in Sec.~\ref{sec:conclusions}.

\section{Theoretical Framework}
\label{sec:theory}
We adopt the standard Press-Schechter (PS) formalism \citep{Press:1973iz, Bond:1990iw}:
\begin{equation}\label{eq:PS_HMF}
    \frac{dn_{\rm CDM}}{d\ln M} = \frac{\bar \rho_{\rm c}\Omega_{\rm m}}{M} f(\nu_c)\, \nu_c \left| \frac{d\ln \sigma (M, z)}{d\ln M}\right|\,.
\end{equation}
Here, $dn_{\rm CDM}/d\ln M$ is the number density of haloes per mass scale interval, and $\bar\rho_{\rm c}$ is the critical density. The quantity $\sigma$, defined as 
\begin{equation}\label{eq:sigma}
    \sigma^2(M) = \int_0^{\infty} \frac{k^2 dk}{2\pi^2}\, W^2(kR)\, P_{\rm m}(k)\,,
\end{equation}
denotes the variance of the linear matter density field smoothed on a comoving scale $R$, with $P_{\rm m}(k)$ the linear matter power spectrum. The peak height $\nu_c \equiv \delta_c / \sigma$ characterises the rareness of a fluctuation, where $\delta_c$ is the critical overdensity for collapse.
Two window functions $W$ are commonly used in the literature. The top-hat filter,
\begin{equation}\label{eq:tophat}
    W_{\rm TH}(kR) = \frac{3}{(kR)^3} \Big[\sin(kR) - kR\cos(kR)\Big]\,,
\end{equation}
is the most physically motivated choice: it corresponds in real space to a uniform sphere of radius $R$, so that the enclosed mass is simply $M = 4\pi \bar\rho_{\rm m} R^3/3$. However, for power spectra with a sharp small-scale suppression such as those arising in warm dark matter cosmologies, the top-hat filter mixes modes above and below the suppression scale, leading to an overestimate of halo abundances at the affected mass scales \citep{Schneider:2013ria, Schneider:2014rda}. In such cases, the sharp-$k$ filter
\begin{equation}\label{eq:sharpk}
    W_{\rm SK}(kR) = \Theta(1 - kR)
\end{equation}
provides a cleaner separation of modes by construction, acting as a hard cutoff in $k$-space. The cost is that the sharp-$k$ filter has no simple real-space counterpart, breaking the trivial relation between the smoothing scale $R$ and the enclosed mass. \citet{Schneider:2013ria, Schneider:2014rda} account for this by introducing a free parameter $c_R$ in the mass--radius relation,
\begin{equation}\label{eq:M2R}
    M = \frac{4\pi}{3}\, \bar\rho_{\rm m}\, (c_RR)^3\,,
\end{equation}
which can be calibrated against simulations. 

The PS multiplicity function reads
\begin{equation}
    f_{\rm PS}(\nu_c) = \sqrt{\frac{2}{\pi}}\, e^{-\nu_c^2/2}\,,
\end{equation}
and gives the fraction of mass elements that have collapsed into haloes above a given mass threshold.

While the PS formalism captures the essential physics, it relies on the assumption of spherical collapse, which implies a constant critical density threshold $\delta_c$ for all masses. \citet{Sheth:1999mn} and \citet{Sheth:2001dp} generalised this work to ``ellipsoidal collapse'' which implies the Sheth-Tormen (ST) multiplicity function
\begin{equation}
    f_{\rm ST}(\nu_c) = A\sqrt{\frac{2q}{\pi}}\left(1 + (q\nu_c^2)^{-p}\right)
    e^{-q\nu_c^2/2}\,.
\end{equation}
The standard parameters providing the best fit to simulations are $p \approx 0.3$, $q \approx 0.707$, and normalization $A \approx 0.322$ \citep{Sheth:1999mn}.

\subsection{The DDM Halo Mass Function}\label{sec:naive}

\begin{figure}
    \centering
    \includegraphics[scale=0.55]{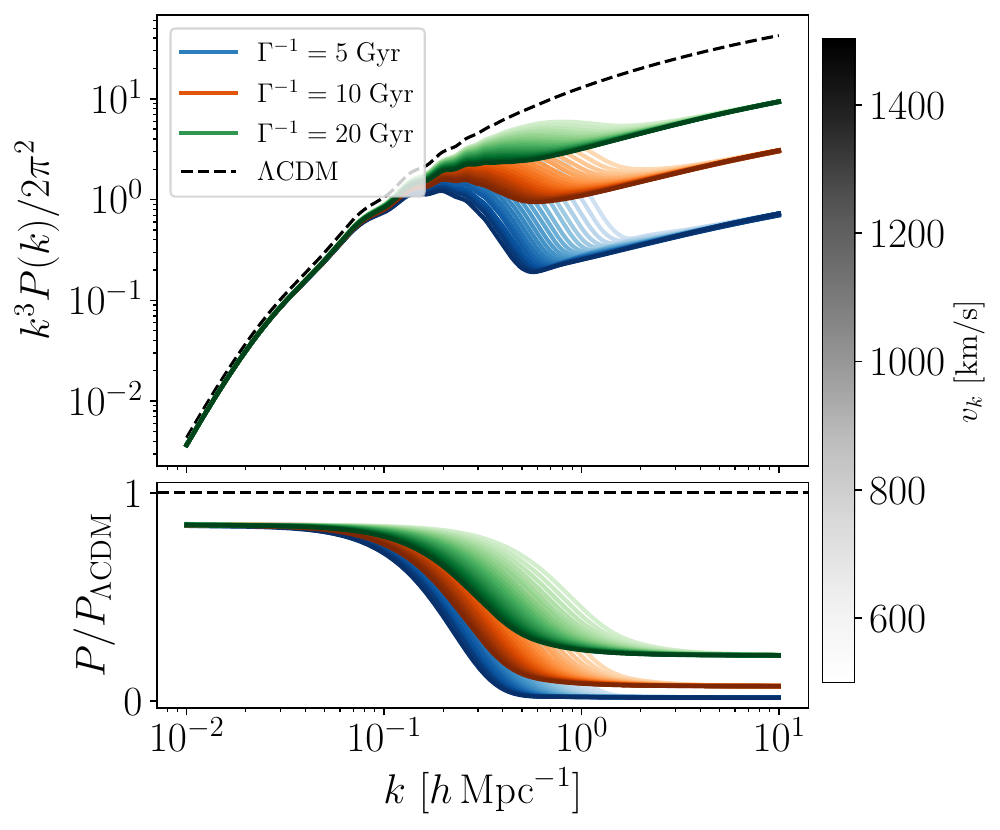}
    \caption{Matter power spectra for various DDM models. \textit{Top panel}: dimensionless power spectrum $k^3 P(k)/2\pi^2$ for $\Lambda$CDM (dashed black) and three DDM lifetimes (distinguished by color), with the velocity kick $v_k$ varying within each lifetime as indicated by the colorbar shading. \textit{Bottom panel}: ratio $P/P_{\Lambda\mathrm{CDM}}$, highlighting the small-scale power suppression induced by the DDM decay.} 
    \label{fig:mpk}
\end{figure}

\begin{figure*}
    \centering
    \includegraphics[scale=0.7]{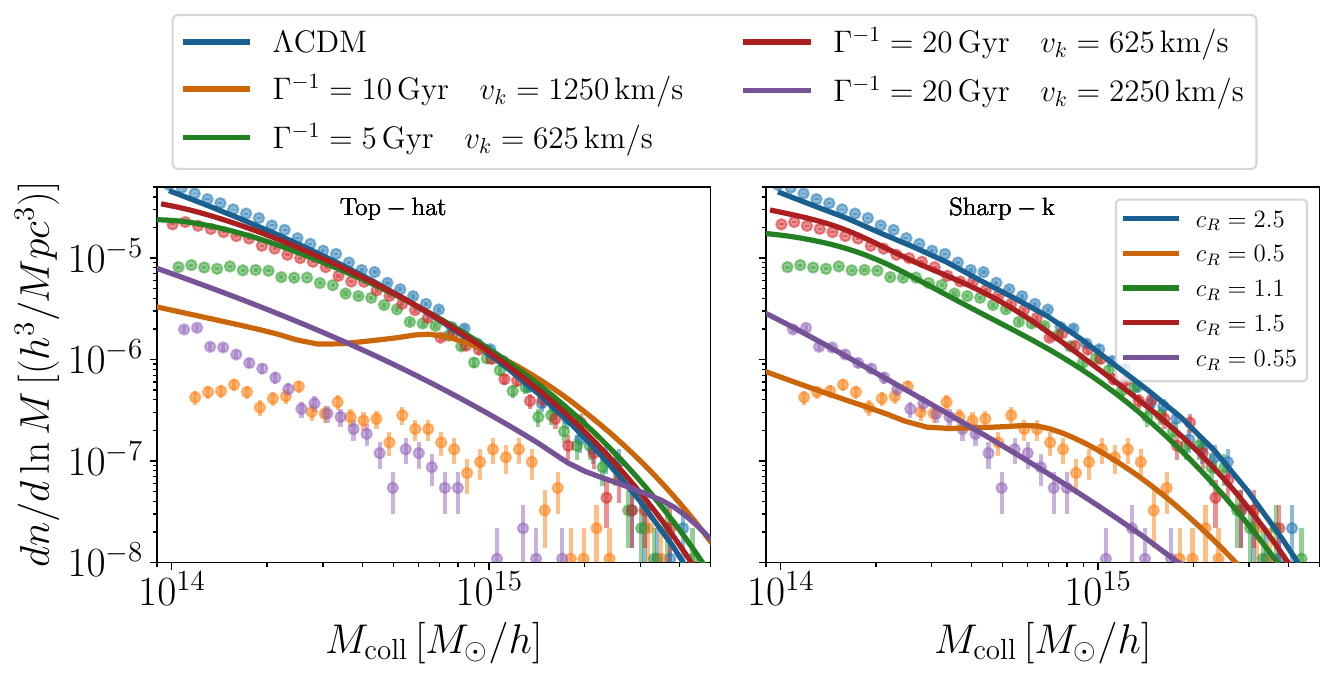}
    \caption{Naive estimation of the halo mass function injecting the DDM linear power spectrum in \eqref{eq:hmf_ps_final} using a top-hat filter (left panel) and a sharp-$k$ filter (right panel). }
    \label{fig:hmf_naive}
\end{figure*}

In DDM, the continuous decay of parent particles and the subsequent creation of energetic daughter particles causes the halo to lose mass, which modifies the mass function. The initial Lagrangian mass of a perturbation, $M_0$, which stays constant in EdS, here decreases and reaches the collapsed mass $M_{\rm coll}$; this mass evolution is derived in Sec.~\ref{sec:DDM Spherical Collapse}. The variance of the density field $\sigma^2$ on the RHS of Eq.~\eqref{eq:PS_HMF} is evaluated using the linear power spectrum at the Lagrangian scale enclosing the mass $M_0$. To map the abundance of these perturbations to the observed halo masses, we introduce a Jacobian transformation $d\ln M_0 / d\ln M_{\rm coll}$. The modified PS HMF for DDM hence reads:
\begin{multline}\label{eq:hmf_ps_final}
    \frac{dn}{d\ln M_{\rm coll}} =
    \frac{\bar{\rho}_{\rm c} \Omega_{m}}{M_0} f(\nu_c) \nu_c \left| \frac{d\ln \sigma}{d\ln M_0}\right| \frac{d\ln M_0}{d\ln M_{\rm coll}}\,.
\end{multline}
Note that the pre-factor $\bar{\rho}_{\rm c} \Omega_{m} / M_0$ is left unchanged because the PS patch counting is performed in Lagrangian space on the initial density field.

To evaluate the HMF, one could straightforwardly inject the DDM linear power spectrum at $z=0$ into eq.~\eqref{eq:sigma} to compute $\sigma$, and then apply the standard PS formalism via eq.~\eqref{eq:hmf_ps_final} with the standard $\Lambda$CDM constant critical density value. As illustrated in Fig.~\ref{fig:mpk}, the DDM power spectrum exhibits a suppression of small-scale power relative to $\Lambda$CDM, a direct consequence of the free-streaming of daughter particles. The amplitude of the suppression is primarily controlled by the decay rate $\Gamma$: a shorter lifetime leaves more time for free-streaming to wash out small-scale structure, producing stronger suppression. The kick velocity $v_k$, on the other hand, sets the scale at which the suppression kicks in.

However, this naive approach fails to reproduce the HMF measured in simulations, as illustrated in Fig.~\ref{fig:hmf_naive} for four DDM models. In the left panel, using the top-hat filter defined in Eq.~\eqref{eq:tophat}, we find a dramatic overestimation of halo abundances across all masses for models with large velocity kicks $v_k = 1250$~km/s in orange and $v_k = 2250$~km/s in purple. As already mentioned, this behaviour is well known in the context of warm dark matter: the top-hat filter in real space has a very broad kernel in $k$-space and cannot capture the sharp small-scale suppression in $P_{\rm m}(k)$, and is typically resolved by adopting a sharp-$k$ filter defined in Eq.~\eqref{eq:sharpk}. The similar small-scale power suppression for both DDM and WDM therefore motivates the same approach. 
In the right panel, each model is shown with its rough best-fit value of $c_R$ (defined in eq.~\eqref{eq:M2R}), ranging from $c_R \approx 0.5$ for the largest kick velocity to $c_R \approx 2.5$ for $\Lambda$CDM. While the high kick velocity models ($v_k = 1250$ and $2250$ km/s) can be reasonably well reproduced, no single value of $c_R$ provides a satisfactory fit for the low kick velocity models ($v_k = 625$~km/s, red and green). 

This suggests that no single standard window function is able to simultaneously reproduce all DDM models when using the DDM linear power spectrum. One could in principle search for a DDM-specific window function, for instance by promoting the sharp-$k$ parameter $c_R$ to a function of the DDM parameters. Such an approach would inevitably introduce model-dependent fitting parameters with no clear physical motivation.

Instead, we adopt a different strategy. We keep the top-hat window function, which is physically motivated by the relation between a Lagrangian mass scale and a comoving smoothing radius $R$, see eq.~\eqref{eq:sigma}, and we use the $\Lambda$CDM linear power spectrum to compute $\sigma(M)$. The DDM physics is then injected directly into the spherical collapse model by accounting for the mass loss induced by the decay of dark matter particles. Mass loss leads to a mass-dependent critical density which we inject directly in $f(\nu_c)$. Using the constant-barrier multiplicity function $f(\nu_c)$ (PS or ST) with a mass-dependent $\nu_c(M) = \delta_c(M)/\sigma(M)$ is an approximation whose accuracy cannot be guaranteed analytically. We adopt it as the simplest extension and validate the full prediction directly against N-body simulations in Sec.~\ref{sec:validation}.

The mass loss is ultimately the same physical effect as the small-scale power suppression in the DDM linear power spectrum --- both arise from the velocity kicks inherited to daughter particles --- but treating it at the level of the collapse dynamics allows for a parameter-free, physically motivated framework, as we now describe.

\subsection{DDM Spherical Collapse}\label{sec:DDM Spherical Collapse}
A spherical overdensity collapses when its linearly extrapolated density contrast reaches a critical threshold $\delta_c$ (see e.g.~\cite{Desjacques:2016bnm} for a review). The key difference in the DDM collapse model is that the decay induces a continuous loss of mass, which softens the gravitational potential and delays the collapse. This delay translates into a higher critical density threshold compared to EdS, making halo formation harder and ultimately suppressing the HMF. 

To compute $\delta_c$, we use the spherical collapse model; a spherical top-hat overdensity in which the enclosed mass evolves due to DDM decay. The key quantities are the parent mass $M_{\rm p}$, the accumulated daughter mass $M_{\rm d}$, and the gravitating mass $M_{\rm grav}$ that physically enters the equation of motion of a halo shell. We derive each of these in turn.

For a homogeneous halo of radius $R(t)$ and initial mass $M_{0}$, the equation of motion for the shell radius can be written in terms of the dimensionless variables $\tilde R = R/R_{\rm ta}$ and $\tilde t = t/t_{\rm ta}$, where the subscript $\rm ta$ stands for ``turn-around'', see appendix~\ref{app:EdS}, 
\begin{equation}\label{eq:EOM}
    \tilde R'' = -\frac{\pi^2}{8 \tilde R^2}\frac{M_{\rm grav}}{M_0}\,,
\end{equation}
where the prime stands for derivatives w.r.t. $\tilde t$. The key difference of the DDM case with respect to standard LCDM is that the gravitating mass is now time-dependent, i.e. the parent mass decays and part of the daughter mass escapes the halo.  

Note that Eq.~\eqref{eq:EOM} contains no explicit cosmological-constant term. Since $\rho_\Lambda$ is constant while the interior matter density grows during collapse, $\rho_\Lambda$ remains negligible compared to self-gravity in the collapse dynamics and shifts only the turn-around epoch at the percent level \citep{Percival:2005vm}.

The parent mass evolves due to decay as
\begin{equation}\label{eq:dotMp}
    M_{\rm p}' = -\tilde \Gamma M_{\rm p} \,,
\end{equation}
where $\tilde \Gamma = t_{\rm ta}\Gamma$ is the dimensionless decay rate. From \eqref{eq:dotMp}, one can derive the bound rate of daughter mass produced. Defining 
\begin{equation}
\epsilon = \frac {v_k/c}{1+v_k/c} \,,
\end{equation}
where $c$ is the speed of light and $v_k$ is the kick velocity inherited by the daughter particle, we can show that 
\begin{equation}\label{eq:dotMd}
    M'_{\rm d} = f_{\rm bound} \sqrt{1-2\epsilon}\, \tilde \Gamma M_{\rm p}(\tilde t)\,,
\end{equation}
where $f_{\rm bound}$, derived below, is the fraction of newly produced daughters that remain gravitationally bound, so that $M_d$ denotes the bound daughter mass. 

The gravitating mass $M_{\rm grav}$ entering Eq.~\eqref{eq:EOM} is, strictly speaking, the total mass of all particles physically enclosed within the shell at a given time $t$. In the standard $\Lambda$CDM case this is simply $M_0$, but in DDM it is time-dependent and non-trivial to evaluate: produced daughter particles may escape the halo if they are unbound, and even bound daughters may have orbits that cross the shell boundary $R$, spending only part of their time inside it. Tracking each daughter particle individually would require a full N-body treatment. Instead, we derive an approximation for $M_{\rm grav}$ as a function of $M_{\rm p}$, $M_{\rm d}$, and two population fractions: $f_{\rm bound}$, the fraction of daughters that are energetically bound to the halo, and $f_{\rm in}$, the fraction whose orbits remain fully inside $R$. We now compute these fractions analytically.

\paragraph{Harmonic potential of the uniform sphere.}
Within the spherical collapse model, the collapsing overdensity is treated as a uniform density sphere throughout the collapse -- this is the standard top-hat approximation. Inside such a sphere of mass 
$M_{\rm grav}$ and radius $R$, the gravitational potential is harmonic:
\begin{equation}\label{eq:Phi_harmonic}
    \Phi(r) = -\frac{G M_{\rm grav}}{2R}\left(3 - \frac{r^2}{R^2}\right) = \frac{1}{2}\omega^2 \left(r^2 - 3R^2\right)\,,
\end{equation}
with $\omega^2 = GM_{\rm grav}/R^3$. 
For a daughter particle treated as a test particle inside the uniform halo, the equation of motion is $\ddot r_i + \omega^2 r_i = 0$ for each Cartesian component. In the special case of constant $\omega$, this is a harmonic oscillator and each Cartesian pair of position and scaled velocity $(r_i,\, \dot{r}_i/\omega)$ traces a circle, with radius $\sqrt{r_i^2 + \dot{r}_i^2/\omega^2}$. Summing over the three components, the full orbit traces a sphere in the six-dimensional phase space with radius
\begin{equation}\label{eq:amplitude}
    \mathcal{A}^2 \equiv \sum_{i=1}^{3} \left(r_i^2 + \frac{\dot r_i^2}{\omega^2}\right) = r^2 + \frac{v^2}{\omega^2}\,,
\end{equation}
which reduces to a conserved quantity when $\omega$ is constant. 

In our more general case, $\omega$ depends on both $R$ and $M_{\rm grav}$, which evolve over time, so $\mathcal{A}^2$ varies and serves instead as a useful instantaneous characterisation of an orbit. For a purely radial orbit, $\mathcal{A}$ coincides with the apocenter, $r_{\rm max} = \mathcal{A}$. For a general orbit, it depends on the angle $\theta$ {(between the orbit direction and the radial direction)} and reads
\begin{equation}\label{eq:incondition}
    r_{\rm max}^2(\theta) = \frac{{\mathcal A^2}}{2} + \sqrt{\left(r^2 - \frac{\mathcal A^2}{2}\right)^2 + \frac{r^2v_k^2}{\omega^2}\cos^2\theta}\,.
\end{equation}
Moreover, written in terms of $\omega$ and $\mathcal{A}$, the total energy reads
\begin{equation}\label{eq:energy}
    E = \frac{1}{2}\omega^2\left(\mathcal{A}^2 - 3R^2\right)\,.
\end{equation}
If the total energy is positive, the particle is unbound. Hence, we can partition the daughter particles into three distinct populations:
\begin{enumerate}
  \item $r_{\rm max}^2(\theta) < R^2$: The orbit is fully contained within the sphere. These daughter particles always contribute to the gravitating mass of the halo.
  \item $r_{\rm max}^2(\theta) > R^2$ and $\mathcal{A}^2 < 3R^2$: The daughter is energetically bound to the halo but follows an orbit that crosses the shell boundary. A bound daughter whose orbit extends beyond $R$ exits the overdense region, reducing its continuous gravitational contribution to the shell equation and effectively ``puffing'' the halo beyond $R$. 
  \item $\mathcal{A}^2 \geq 3R^2$: The daughter particle is unbound and permanently escapes the host halo. 
\end{enumerate}
After their creation, daughter particles evolve in the time-dependent potential of the collapsing sphere, since both $R$ and $M_{\rm grav}$ vary and therefore $\omega$ is not constant. Consequently, the quantities $\mathcal{A}^2$ and $r_{\rm max}^2(\theta)$ defined above are not exact integrals of motion, but only instantaneous orbital diagnostics evaluated in the potential at the time of decay. In this work we therefore adopt an instantaneous-orbit approximation: each daughter particle is classified using its amplitude $\mathcal{A}^2$ and maximal orbit $r_{\rm max}^2(\theta)$ at the moment of creation, and is subsequently evolved according to its assigned population. This avoids tracking the full distribution of daughter trajectories in the evolving potential, which would be computationally prohibitive. The approximation is best motivated by the fact that the characteristic timescale of spherical collapse is comparable to a single orbital period, over which the phase-space evolution of a daughter particle remains limited. We adopt this approximation throughout this work.

\paragraph{Fraction of bound particles.}
For an isotropic decay, a parent particle at radius $r$ moves with radial velocity $v_{p}(r) = (\dot{R}/R)\,r$ and receives a velocity kick $v_k$ at angle $\theta$ relative to the radial direction. The daughter velocity immediately after the kick is $v_d^2 = v_{p}^2 + v_k^2 + 2v_{p}v_k\cos\theta$. 
Defining the dimensionless variables 
\begin{align}\label{eqs:beta_xi_u}
    \beta = \frac{|\dot R|}{\omega R}\,,\qquad
    \xi = \frac{v_k}{\omega R}\,,\qquad
    u = \frac{r}{R}\,,
\end{align}
and using Eq.~\eqref{eq:amplitude}, the condition $\mathcal{A}^2 < 3 R^2$ translates into a condition on the kick angle, 
\begin{equation}\label{eq:Cr}
    \cos\theta < C_{\rm bound}(u) \equiv \frac{1}{2\beta \xi u} \Big(3-\xi^2 - u^2 \left[1+\beta^2\right]\Big)\,.
\end{equation}
For isotropic decay, the probability for a daughter created at radius $r$ to satisfy $\mathcal{A}^2 < 3 R^2$ is
\begin{equation}\label{eq:Pbound}
    P_{\rm bound}(u) = \begin{cases} 1 & C_{\rm bound} \geq 1\,, \\ \dfrac{1+C_{\rm bound}(u)}{2} & -1 \leq C_{\rm bound} < 1\,, \\ 0 & C_{\rm bound} < -1\,, \end{cases}
\end{equation}
where the conditions may be written in terms of $u$
\begin{align}
    u &\leq u_1 = \frac{- \beta\xi + \sqrt{3(1+\beta^2) - \xi^2}}{1+\beta^2}\\
    u &\leq u_2 = \frac{\beta\xi + \sqrt{3(1+\beta^2) - \xi^2}}{1+\beta^2}\,.
\end{align}
The corresponding volume-averaged fraction can be computed analytically. It simply reads $f_{\rm bound}=1$ if $u_1 \geq 1$, else
\begin{align}\label{eq:fbound}
    f_{\rm bound} = u_1^3 + 3\left[\frac{u^3}{6} + \frac{3-\xi^2}{8\beta\xi}u^2 - \frac{1+\beta^2}{16\beta\xi }u^4 \right]_{u_1}^{\min(1, u_2)}\,.
\end{align}

\paragraph{Fraction of fully contained particles.}
To evaluate the fraction of daughters whose orbits remain fully contained inside the halo, $f_{\rm in}$, we note that the parent bulk flow $v_p=(\dot R/R)\,r$ is homologous: it scales the daughter's radius and the shell radius $R$ by the same factor, leaving the ratio $r_{\rm max}/R$ invariant. Only the peculiar velocity imparted by the kick alters this ratio. We therefore evaluate the instantaneous apocentre retaining only $v^2=v_k^2$ in Eq.~\eqref{eq:incondition} and compare it with the instantaneous halo radius $R$. In terms of $u$ and $\xi$, the condition $r_{\rm max}^2<R^2$ becomes 
\begin{equation}
    \cos^2\theta <
    \frac{(1-\xi^2)(1-u^2)}{u^2\xi^2}\,.
\end{equation}
The fraction of kick directions satisfying this inequality is
\begin{equation}
    P_{\rm in}(u) = \min\left[ 1, \frac{1}{u\xi} \sqrt{(1-\xi^2)(1-u^2)} \right]\,,
\end{equation}
for $\xi\leq1$, and $P_{\rm in}=0$ for $\xi>1$. Averaging over the uniform volume distribution of decay positions gives
\begin{equation}\label{eq:fin}
    f_{\rm in}
    =
    \begin{cases}
    \sqrt{1-\xi^2}\,, & \xi\leq1\,,\\
    0\,, & \xi>1\,.
    \end{cases}
\end{equation}
This expression is $\beta$-independent, in contrast to $f_{\rm bound}$: the containment ratio is invariant under the bulk flow, whereas the binding energy that defined $f_{\rm bound}$ receives a genuine contribution from the bulk motion. The instantaneous evaluation neglects a correction of order $\beta$, which vanishes at turnaround and is largest near the initial conditions and near collapse. Neglecting it allows $f_{\rm in}$ to formally exceed $f_{\rm bound}$ in this large-$\beta$ regime, since $f_{\rm in}$ is bulk-flow-independent while $f_{\rm bound}$ is suppressed by the bulk motion. Near the initial conditions this is harmless because negligible daughter mass has yet been produced ($M_{\rm d}\to0$); near collapse it is confined to a thin region that vanishes as $\xi\to0$ at $R\to0$, where $f_{\rm in}=f_{\rm bound}=1$.

\paragraph{Gravitating Mass.}
By construction, $f_{\rm bound}$ counts all bound daughter particles (populations 1 and 2), while $f_{\rm in}$ counts only those whose orbits remain fully inside the sphere (population 1). By Newton's shell theorem, only the mass physically enclosed within $R$ contributes to the gravitational force on the shell. The three populations contribute as follows: population 1 (fraction $f_{\rm in}$) never crosses the shell and always contributes; population 2 (fraction $f_{\rm bound}-f_{\rm in}$) is bound but its orbit crosses $R$; population 3 (fraction $1-f_{\rm bound}$) is unbound and never contributes.

For population 2, we adopt the following effective prescription. When daughter particles are rare ($M_d/(M_d+M_p) \ll 1$), the halo is parent-dominated and population 2 daughter particles exit $R$ before collapse completes and they do not contribute. When daughter particles dominate, they define the halo themselves and all bound daughter particles contribute. Interpolating with the fraction of daughter particle as the natural parameter gives
\begin{equation}\label{eq:Mgrav}
    M_{\rm grav} = M_p + \left[\frac{f_{\rm in}}{f_{\rm bound}} + \frac{M_d}{M_{d} + M_{p}} \left(1 - \frac{f_{\rm in}}{f_{\rm bound}}\right)\right]\,M_d\,.
\end{equation}
When $v_k = 0$, $f_{\rm in} = f_{\rm bound} = 1$ and the standard top-hat is recovered. This gradual transfer of population 2 into the gravitating mass as daughter particles build up is the \textit{halo puffing} effect: the daughter distribution progressively extends the effective halo out to $\sqrt{3}R$.

Finally, we define the collapsed mass as the gravitating mass evaluated at the collapsed time $t_{\rm coll}$, i.e. when $R \rightarrow 0$.

\subsection{Initial conditions.}
We set initial conditions for the system \eqref{eq:EOM}, \eqref{eq:dotMp} and \eqref{eq:dotMd} by considering a spherical top-hat perturbation of total mass $M_0$ with initial overdensity $\delta_0$. The initial time $t_0$ is chosen to lie deep in matter domination and before any significant decay, so that the EdS approximation holds exactly
\begin{equation}\label{eq:EdS}
    H(t_0) = \frac{2}{3t_0}\,,\qquad 
    \bar\rho_{\rm m}(t_0) = \frac{1}{6\pi G t_0^2}\,.
\end{equation}
The initial radius and mass compartments follow directly,
\begin{align}
    R(t_0) &= \left(\frac{3M_0}{4\pi\bar\rho_{\rm m}(t_0)(1+\delta_0)} 
    \right)^{1/3}\,,\nonumber\\
    M_p(t_0) &= M_0\,,\nonumber\\
    M_d(t_0) &= 0\,.
\end{align}
The initial shell velocity is set using second-order Lagrangian perturbation theory, which corrects the pure Hubble flow for the effect of the local overdensity,
\begin{equation}
    \dot R(t_0) = H(t_0)\,R(t_0)
    \left(1 - \frac{\delta_0}{3} - \frac{2\delta_0^2}{21}\right)\,.
\end{equation}
\smallskip

We adopt EdS initial conditions throughout rather than extracting them from the Boltzmann solver \texttt{CLASS} \citep{Lesgourgues:2011re, Blas:2011rf}. Although \texttt{CLASS} provides a more accurate background at early times, the standard spherical collapse framework is formulated in EdS, and we found that the small residual radiation contributions in the initial conditions introduce corrections that accumulate over the collapse and degrade the accuracy of the solution.

\begin{figure}
    \centering
    \includegraphics[width=1\linewidth]{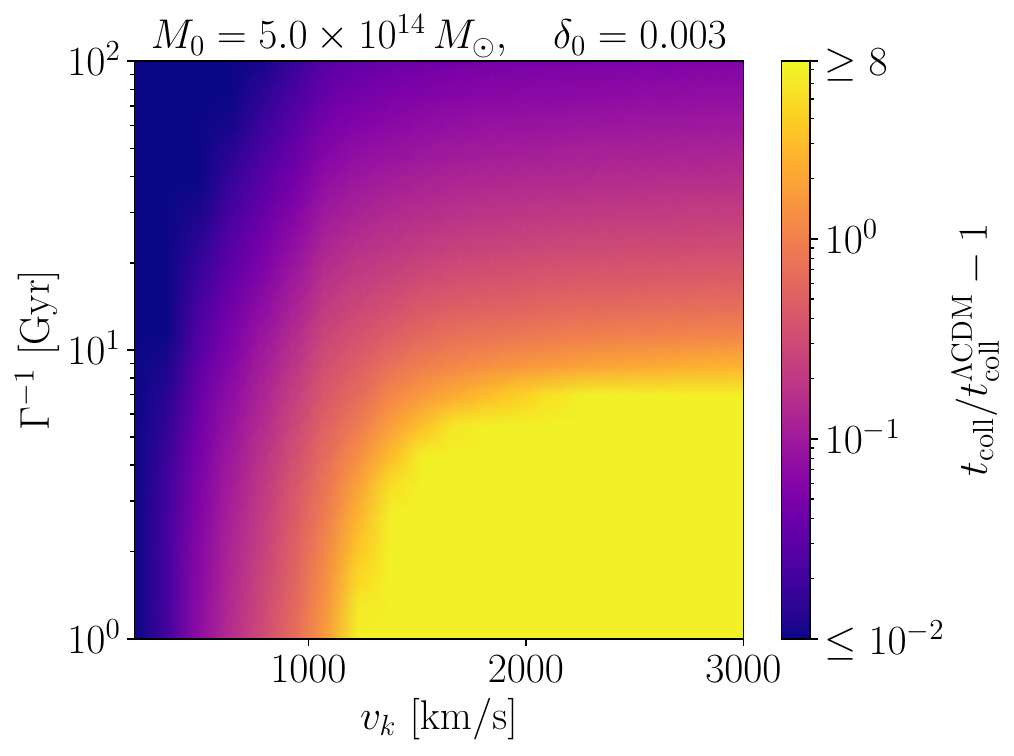}
    \caption{Ratio of the halo collapse time to the $\Lambda$CDM collapse time, $t_{\rm coll}/t_{\rm coll}^{\Lambda{\rm CDM}} - 1$, as a function of the kick velocity $v_k$ and the lifetime $\Gamma^{-1}$, for a top-hat overdensity with initial mass $M_0 =5\times10^{14}\,M_\odot$ and $\delta_0 = 0.003$. DDM effects are negligible at small kick velocities and long lifetimes (ratio $\simeq 1$), while short lifetimes combined with large kicks significantly delay.}
    \label{fig:collapse_time_ratio}
\end{figure}

\subsection{Linear extrapolation and critical density}
\label{sec:linear_extrap}

\begin{figure}
    \centering
    \includegraphics[width=0.95\linewidth]{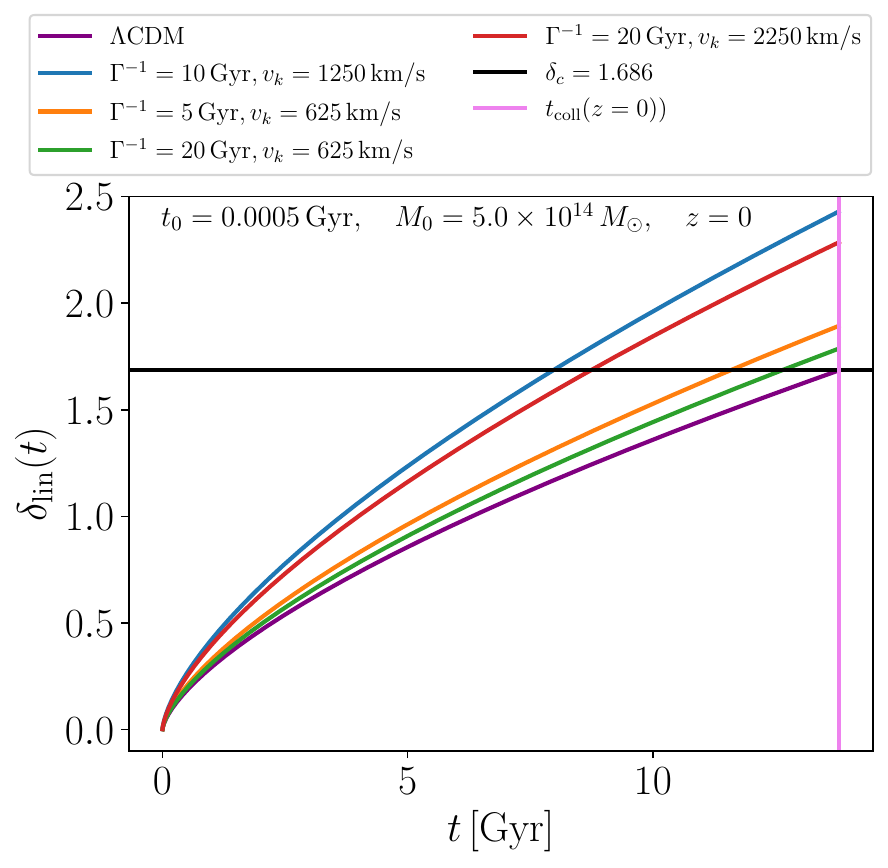}
    \caption{Linear density contrast $\delta_{\rm lin}(t)$ as a function of cosmic time for $\Lambda$CDM and four DDM models, for a top-hat overdensity with initial mass $M_0 = 5\times10^{14}\,M_\odot$ initialised at $t_0 = 5\times10^{-4}\,\mathrm{Gyr}$. The initial overdensity is chosen such that the collapse happens at $z=0$ for each model. The horizontal black line marks the standard $\Lambda$CDM value $\delta_c = 1.686$, and the vertical line indicates the desired collapse time $z = 0$. The $\Lambda$CDM reference is obtained with the same pipeline setting $v_k = 0$ and $\Gamma = 0$.}
    \label{fig:delta_lin}
\end{figure}

\begin{figure*}
    \centering
    \includegraphics[width=1\linewidth]{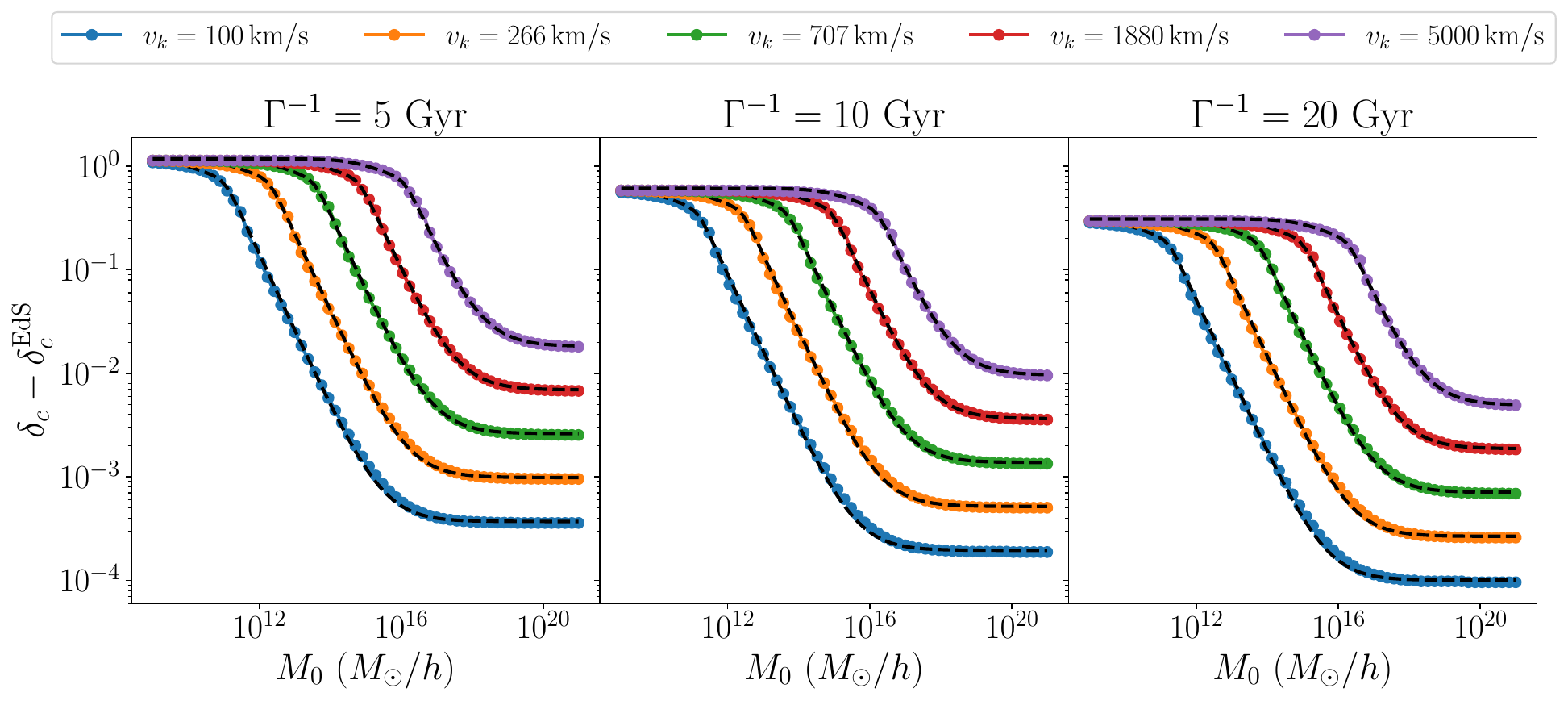}
    \caption{Mass dependence of the collapse threshold $\delta_c$ subtracted by $\delta_c^{\rm EdS}$ for $5$ velocity kicks (colored lines), and $3$ lifetime (panels). At small masses, $\delta_c$ plateaus to a constant value that depends primarily on $\Gamma^{-1}$. At large masses, $\delta_c$ converges to a value close to, but slightly larger than $\delta_c^{\rm EdS}$ for all models.} 
    \label{fig:delta_c}
\end{figure*}

We therefore have three initial conditions: $t_0$, $M_0$, and $\delta_0$. The collapse time is defined as $R(t_{\rm coll}) \to 0$. Fig.~\ref{fig:collapse_time_ratio} illustrates, for a given set of initial conditions, the impact of the DDM parameters on the collapse time normalised by the CDM collapse time $t_{\rm coll}/t_{\rm coll}^{\Lambda{\rm CDM}} - 1$ as a function of $v_k$ and $\Gamma^{-1}$. As expected, the collapse time converges to the $\Lambda$CDM value (dark blue) in the limits $v_k \to 0$ or $\Gamma^{-1} \to \infty$. The delay is maximised for large velocity kicks and short lifetimes (yellow region), where mass loss is most efficient. 

In practice, however, we wish to compute the halo mass function at a target redshift $z_{\rm target}$. Fixing $t_0$, $M_0$ and $z_{\rm target}$ removes one degree of freedom, leaving $\delta_0$ as the only free parameter: it must be tuned such that the nonlinear collapse occurs at exactly $t_{\rm coll}(z_{\rm target})$, defined by $R(t_{\rm coll}) \to 0$. We determine $\delta_0$ via a bisection algorithm, iterating the shell equation of motion~\eqref{eq:EOM} at each step until convergence. 

Once $\delta_0$ is determined for a given $\{t_0, M_0, z_{\rm target}\}$, 
the critical overdensity is obtained by linearly extrapolating the initial overdensity $\delta_0$ forward to the collapse time $t_{\rm coll}(z_{\rm target})$,
\begin{equation}
    \delta_c(t_0, M_0, z_{\rm target}) = \delta_{\rm lin}(t_0, M_0, z_{\rm target})\,.
\end{equation}
Since the DDM physics is already captured by the spherical collapse dynamics, using the DDM growth factor would double-count it. Hence, we adopt the EdS growth factor
\begin{equation}\label{eq:lin_EdS}
    \delta_{\rm lin}(t) = \delta_0\, \frac{D_{\rm EdS}(t)}{D_{\rm EdS}(t_0)}\,, 
    \qquad D_{\rm EdS}(t) \propto t^{2/3}\,,
\end{equation}
so that all DDM physics enters exclusively through the modified collapse time $t_{\rm coll}$. Extrapolating to the CDM collapse time recovers the universal value 
\begin{equation}
\delta_{\rm c}^{\rm EdS} = \frac{3}{5} \left(\frac{3\pi}{2}\right)^{2/3} \approx 1.686\,,    
\end{equation}
while extrapolating to the delayed DDM collapse time yields a larger $\delta_c$, reflecting the fact that mass loss slows the collapse and makes halo formation harder. This procedure is applied independently for each mass.

The linear extrapolation is illustrated in Fig.~\ref{fig:delta_lin}, which shows $\delta_{\rm lin}(t)$ for $\Lambda$CDM (by setting $\Gamma=0$ and $v_k=0$ in our pipeline, demonstrating that it smoothly converges to $\Lambda$CDM) and four DDM models. We use $t_0=0.0005$~Gyr, $M_0=5\times 10^{14}~M_{\odot}$ and $\delta_0$ such that the collapse happens at the chosen redshift $z=0$, indicated as a vertical line. As expected, $\Lambda$CDM converges to the universal value $\delta_{\rm c}^{\rm EdS}$ which is indicated as the horizontal line. All DDM models require a larger initial $\delta_0$ than $\Lambda$CDM to collapse at the same target redshift. Since $\delta_{\rm lin}$ is extrapolated from this larger $\delta_0$, they reach a larger linearly extrapolated value at $t_{\rm coll}$, which defines a model- and halo mass-dependent $\delta_c > \delta_{\rm c}^{\rm EdS}$. We then compute the HMF using Eq.~\ref{eq:hmf_ps_final} and the value of $\delta_c(\Gamma,\epsilon,M_0)$ obtained from the collapse calculation and $M_{\rm coll}(M_0)$ used for the mass mapping.

\section{Critical density Modeling}\label{sec:Critical_density}
In Fig.~\ref{fig:delta_c} we show the deviation of the critical density $\delta_c$ from $\delta_c^{\rm EdS}$, evaluated numerically for three different lifetimes (panels) and five different velocity kicks shown as the colored solid lines. The black dashed lines are obtained with the fitting function introduced below.

The main feature of $\delta_c(M)$ is a mass-dependent transition between two plateau values: a small-mass plateau that depends only on $\Gamma$ -- within each panel, all curves with different $v_k$ converge to the same value -- and a large-mass plateau close to $\delta_c^{\rm EdS}$ that depends mostly on the velocity kick $v_k$. These two limits correspond to physically transparent cases. In the small-mass limit, halos are too small to retain any daughter particles ($f_{\rm in} = f_{\rm bound} = 0$), so the collapse dynamics depend only on the decay rate $\Gamma$ and not on $v_k$. In the large-mass limit, halos are massive enough to retain all daughter particles ($f_{\rm in} = f_{\rm bound} = 1$), the collapse threshold approaches the EdS value, with only a small residual shift due to the energy carried away by dark radiation. We now discuss these two limits in detail.

Computing $\delta_c$ for an arbitrary DDM model requires solving the full system of equations~\eqref{eq:EOM}, \eqref{eq:dotMp}, \eqref{eq:dotMd} together with~\eqref{eq:Mgrav} numerically via the shooting method of Sec.~\ref{sec:linear_extrap}, which must be repeated for each mass $M_0$, redshift, and each set of DDM parameters $(\Gamma, v_k)$. While accurate, this is computationally expensive and provides little physical insight into the parameter dependence. We therefore derive analytical expressions for the two plateau values and provide a fitting formula for the transition. This makes the framework directly usable in cosmological parameter inference without running multiple times the collapse ODE at each likelihood evaluation. We now derive each ingredient in turn.

\subsection{Large-mass limit}
\begin{figure}
    \centering
    \includegraphics[width=1\linewidth]{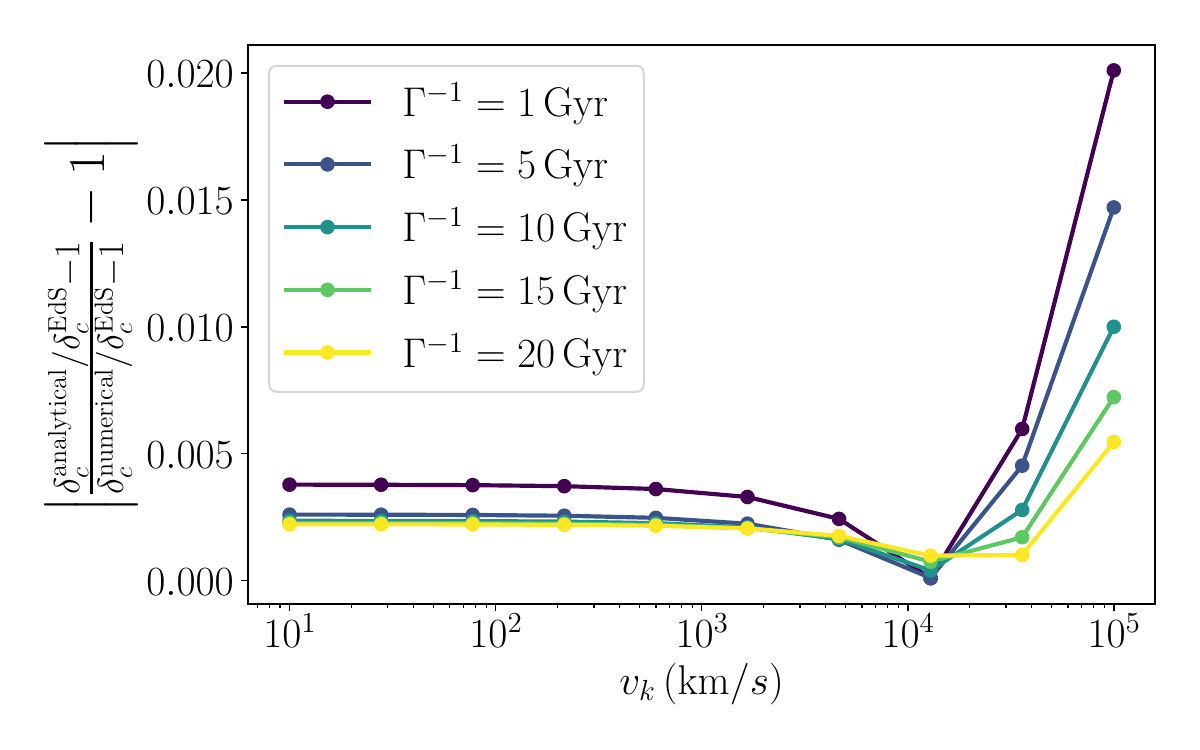}
    \caption{Relative deviation between the large-mass analytical approximation $\delta_c^{\rm analytical}$ of Eq.~\eqref{eq:deltac_large} and the large-mass-limit numerical result $\delta_c^{\rm numerical}$, as a function of the velocity kick $v_k$ for five decay lifetimes as indicated in the legend.}
    \label{fig:analytical_vs_num}
\end{figure}
At large masses, massive halos collapse in a deep potential well such that all daughter particles are retained. The only difference with respect to $\Lambda$CDM is then the small fraction of rest-mass energy carried away by dark radiation. In the limit $\epsilon \ll 1$, which holds for all the models we study since $v_k \ll c$, we can linearize the ODE 

\begin{equation}\label{eq:largemass_ODE}
    \tilde R'' = -\frac{\pi^2}{8 \tilde R^2} \left[ 1+\epsilon \left(e^{-\tilde \Gamma \tilde t} - 1\right) +\mathcal O(\epsilon^2)\right]\,.
\end{equation}
Note that, for the EdS cycloid, $t_{\rm ta} = t_{\rm coll}/2$, so the zeroth-order dimensionless collapse time is $\tilde t_{\rm coll}^{\rm EdS} = 2$. Consequently, at this order in $\epsilon$, and apart from the overall dependence on $\epsilon$, the critical threshold depends on $\Gamma$ and $z$ only through the combination $\tilde\Gamma = \Gamma\,t_{\rm coll}(z)/2$: two models with different decay rates and observation redshifts but identical $\tilde\Gamma$ have the same large-mass correction to $\delta_c$.

At zeroth order, we recover the EdS limit summarised in appendix~\ref{app:EdS} leading to $\delta_c^{\rm EdS}$. The solution of eq.~\eqref{eq:largemass_ODE} can be written in the form $\tilde R = \tilde R_{\rm EdS} + \epsilon \tilde R_1$. Substituting in Eq.~\eqref{eq:largemass_ODE} and linearising in $\epsilon$, we obtain at first order the $\epsilon$-free ODE
\begin{align}\label{eq:largemass_ODE2}
    \tilde R_1'' - \frac{2\pi^2}{8\tilde R^3_{\rm EdS}}\tilde R_1 
    = -\frac{\pi^2}{8\tilde R_{\rm EdS}^2}
    \left(e^{-\tilde\Gamma\tilde t} - 1\right)\,,
\end{align}
whose only dependence on the decay rate $\tilde \Gamma$ and the observation redshift enters solely through the combination $\tilde\Gamma = \Gamma\,t_{\rm coll}(z)/2$.

The resolution of this differential equation is discussed in Appendix~\ref{app:largemasslimit}. The particular solution is dominant and reads
\begin{multline}\label{eq:part}
    R^{\rm part}_1(\theta) = \frac{1}{2}  \frac{\sin\theta}{1-\cos(\theta)} \int_0^\theta d\theta'\,  \frac{\sin\theta'}{(1-\cos(\theta'))^2}\\
   \left(1 - e^{-\tilde \Gamma t(\tilde \theta')}\right) \left[I(\theta) - I(\theta')\right]\,,
\end{multline}
where
\begin{equation}\label{eq:I}
    I(\theta) = \sin\theta-3 \theta  + 4 \tan \frac{\theta}{2}\,.
\end{equation}

We now determine the collapse time shift by asymptotic matching near the EdS singularity (see appendix~\ref{app:largemasslimit} for details). Writing $\tau_{\rm EdS} \equiv 2-\tilde t$, the perturbed solution approaches
\begin{equation}\label{eq:nongeneric_form}
    \tilde R \approx \frac{(6\pi)^{2/3}}{4}\,\tau_{\rm EdS}^{2/3} 
    - \frac{\epsilon\, J(\tilde \Gamma)}{(6\pi)^{1/3}}\,\tau_{\rm EdS}^{-1/3}\,,
\end{equation}
where
\begin{equation}\label{eq:J}
    J(\tilde \Gamma) = -\int_0^{2\pi} d\theta'\,
    \frac{\sin\theta'\left(1-e^{-\tilde \Gamma \tilde t(\theta')}\right)}
    {(1-\cos\theta')^2}\left[6\pi + I(\theta')\right]\,.
\end{equation}
Near any spherical collapse, the universal form $\tilde R \propto (\tilde t_{\rm coll} - \tilde t)^{2/3}$ holds. Since $\tilde t_{\rm coll} = 2 + \delta\tilde t$, we have $\tilde t_{\rm coll} - \tilde t = \tau_{\rm EdS} + \delta\tilde t$. 
Expanding the universal form to first order in $\delta\tilde t$ and matching the $\tau_{\rm EdS}^{-1/3}$ coefficient with eq.~\eqref{eq:nongeneric_form} gives
\begin{equation}\label{eq:deltat}
    \delta\tilde t = -\frac{\epsilon\, J(\tilde\Gamma)}{\pi}\,.
\end{equation}
Note that $J < 0$, confirming that the collapse is delayed relative to $\Lambda$CDM. 
The linear extrapolation to the collapse time then gives
\begin{equation}\label{eq:deltac_large}
    \delta_c^{\rm large} = \frac{3}{20}\!\left(6\pi\,\tilde t_{\rm coll}\right)^{2/3}
    \approx \delta_c^{\rm EdS}\!\left(1 - \frac{\epsilon\,J(\tilde\Gamma)}{3\pi}\right).
\end{equation}

In Fig.~\ref{fig:analytical_vs_num}, we show the relative error of the analytical formula \eqref{eq:deltac_large} compared to the numerical result in the large-mass limit as a function of the velocity kick and for lifetimes going from $1$~Gyr and $20$~Gyr. The relative error remains $<1\%$ for velocities $\lesssim10^5$~km$/s$.

\subsection{Small-mass and massless limit}
\begin{figure}
    \centering
    \includegraphics[width=1\linewidth]{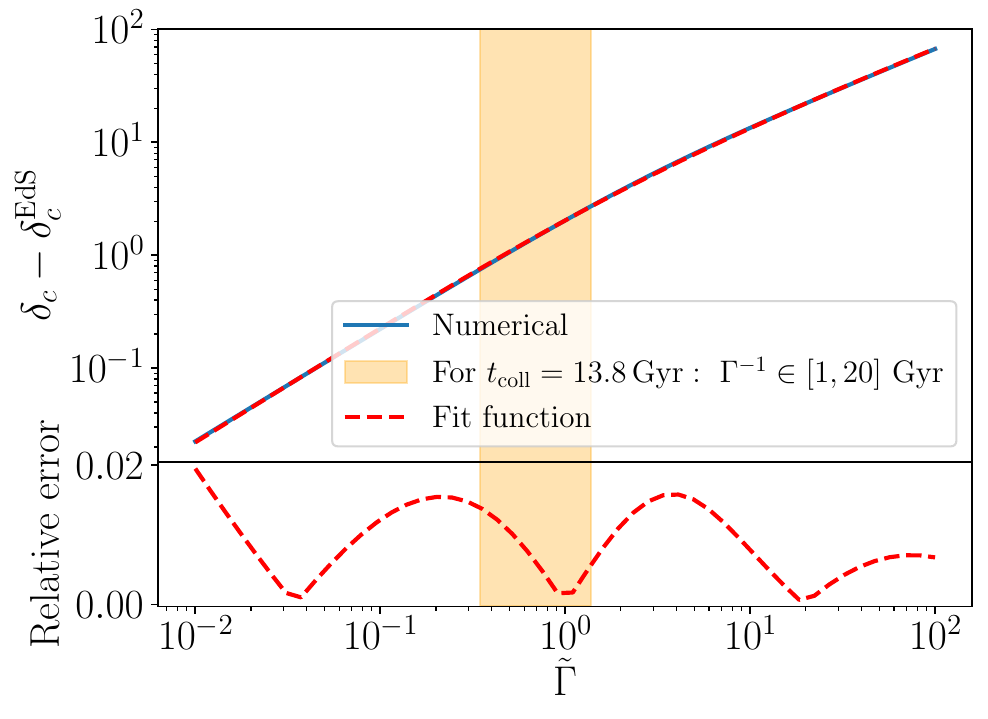}
    \caption{Small-mass plateau of the collapse threshold as a function of $\tilde \Gamma$. \textit{Top panel}: DDM excess $\delta_c - \delta_c^{\rm EdS}$ computed numerically (blue points) and the fitting formula (red dashed, Eq.~\eqref{eq:deltac_plateau}). \textit{Bottom panel}: Relative error between the numerical result and the fit. The orange band represents the relevant range of $\tilde \Gamma$ for a collapse time today.}
    \label{fig:SmallMass}
\end{figure}

In the small-mass limit, halos are so small that the kick velocity $v_k$ always exceeds the escape velocity, so all daughter particles escape immediately and the collapse is driven purely by the decaying parent mass $M_p = M_0\,e^{-\Gamma t}$. The kick velocity therefore drops out entirely. This explains why all models sharing the lifetimes converge to the same small-mass plateau regardless of $v_k$, see Fig.~\ref{fig:delta_c}. 

This small-mass limit is physically equivalent to the case where the parent decays entirely into massless dark radiation, with no massive daughter. In that scenario, $f_{\rm bound} = f_{\rm in} = 0$ identically at all masses since there are no massive daughters to retain. The collapse is therefore also driven by the decaying parent mass $M_{\rm p} = M_0 e^{-\Gamma t}$, giving a critical density $\delta_c$ that is mass-independent. This is precisely the small-mass plateau $\delta_c^{\rm small}(\tilde\Gamma)$, where the same condition $f_{\rm bound} = f_{\rm in} = 0$ holds because all massive daughters escape. The HMF for a purely dark-radiation decay can therefore be recovered from our framework by simply setting $\delta_c = \delta_c^{\rm small}(\tilde\Gamma)$.

In this limit, the dynamics reduces to a single ODE,
\begin{equation}\label{eq:smallmassODE}
    \tilde R'' = -\frac{\pi^2}{8 \tilde R^2}e^{-\tilde \Gamma \tilde t} \,.
\end{equation}
Similarly to \eqref{eq:largemass_ODE2}, this equation contains a symmetry between the particle lifetime $\Gamma^{-1}$ and the observational redshift, which enter only through $\tilde \Gamma$. 

To our knowledge, there is no known analytical solution to \eqref{eq:smallmassODE}. Hence, we compute this function numerically via the same shooting method as in the general case, see Section~\ref{sec:linear_extrap}, and find that it is well described by the fitting formula
\begin{equation}\label{eq:deltac_plateau}
    \delta_c^{\rm small}(\tilde \Gamma) = \delta_{\rm c}^{\rm EdS} + A \tilde \Gamma^{\beta}\ln(1 + \tilde \Gamma)^{1-\gamma}\,.
\end{equation}
Fitting the three parameters, we obtain $A=2.3824$, $\beta=0.5818$, and $\gamma=0.5642$. Note that $\beta$ and $\gamma$ are very close, but imposing $\beta=\gamma$ degrades the fit so that we choose to keep them free. However, we note that the case $\beta=\gamma$ can be used in the limit $\tilde \Gamma \ll 1$, as \eqref{eq:smallmassODE} can then be solved perturbatively, similarly to the large-mass limit, and provide an analytical formula for $A$, leaving only one free parameter to be fitted.

For simplicity, we stick to \eqref{eq:deltac_plateau}. The best fit and the numerical results are shown in Fig.~\ref{fig:SmallMass}. The relative error between the fit and the numerical results, shown in the bottom panel, is better than $1.5\%$ over the range $\Gamma^{-1} \in [1, 20]$ at redshifts near $z=0$, indicated by the orange band. If one imposes $\beta=\gamma$ (best fit: $A=2.2940$ and $\beta=0.6109$), the relative error is roughly multiplied by 2, reaching $1.5$--$3\%$

\subsection{Intermediate mass}
In the previous section, we derived the analytical large- and small-mass plateaus of $\delta_c(M)$, visible for all models in Fig.~\ref{fig:delta_c}. We now fit the transition between these two plateaus in order to predict $\delta_c(M)$ directly from the DDM parameters without numerically solving the collapse ODE for each mass and model. We find that the critical density is well described by
\begin{equation}\label{eq:fitting_function}
    \delta^{\rm fit}_c(M) = \delta_c^{\rm large} + 
    \frac{\delta_c^{\rm small} - \delta_c^{\rm large}}
    {\left[\left(1 + \dfrac{M}{M_1}\right)
    \left(1 + \left(\dfrac{M}{M_2}\right)^4\right)\right]^\nu}\,,
\end{equation}
where $\delta_c^{\rm large}$ and $\delta_c^{\rm small}$ are the analytical limits derived in Eqs.~\eqref{eq:deltac_large} and \eqref{eq:deltac_plateau}. The function recovers the correct limits by construction: as $M \to 0$, the denominator tends to unity and $\delta_c^{\rm fit} \to \delta_c^{\rm small}$, while as $M \to \infty$, the denominator diverges and $\delta_c^{\rm fit} \to \delta_c^{\rm large}$. The product of two generalised Hill functions in the denominator captures the asymmetric shape of the transition: the first factor involving $M/M_1$ describes the gradual onset of the transition at low masses, while the second factor, with its steeper $M^4$ dependence, captures the rapid drop at higher masses. By calibrating the shape parameters against the full grid of numerical solutions, we find that the exponent $\nu = 0.1484$ and the mass ratio $M_2/M_1 = 10^{1.3795} \approx 24$ are universal across all DDM models considered, leaving $M_1$ as the single free parameter per model. 

A further simplification arises when $M_1$ is expressed in terms of the DDM parameters. Fitting $M_1$ across the full grid of $(\Gamma, v_k)$ models, we find the simple power-law relation
\begin{equation}\label{eq:M1_fit}
    M_1 = B v_k^3 \tilde\Gamma^{-0.5} t_{\rm ta}\,,
\end{equation}
with best-fit $\log_{10} B = 3.017$, accurate to $10\%$ across the parameter grid shown in Fig.~\ref{fig:delta_c} and for both redshifts $z=0$ and $z=1.083$. Fitting all exponents freely recovers $M_1 \propto v_k^{2.97}\, t_{\rm ta}^{1.01}$, confirming the $v_k^3$ and $t_{\rm ta}$ scalings derived below across the two redshifts. 

The $v_k^3$ scaling can be understood analytically from the condition $\xi_{\rm ta} = 1$ (see Eq.~\eqref{eqs:beta_xi_u}), i.e.\ the mass scale at which the kick velocity equals the orbital velocity $\omega R$ evaluated at turnaround, from which we obtain
\begin{equation}\label{eq:M1_analytic}
    M_1 \approx \frac{2\sqrt{2}}{\pi G}v_k^3 t_{\rm ta}\,,
\end{equation}
where we have used $M_{\rm grav} \sim M_1$. The $\tilde\Gamma$ dependence arises because the gravitational mass at turnaround $M_{\rm grav}(t_{\rm ta}) \leq M_0$ depends on how much mass has decayed and escaped by then. Since the transition lies outside the strict large- and small-mass limits, we keep the remaining dependence as the empirical scaling $\tilde\Gamma^{-1/2}$ measured from the numerical solutions\footnote{a free fit prefers a slightly smaller exponent ($\approx 0.42$), consistent with $-1/2$ within the scatter given our two-redshift baseline.}.

\begin{figure*}
    \centering
    \includegraphics[width=1\linewidth]{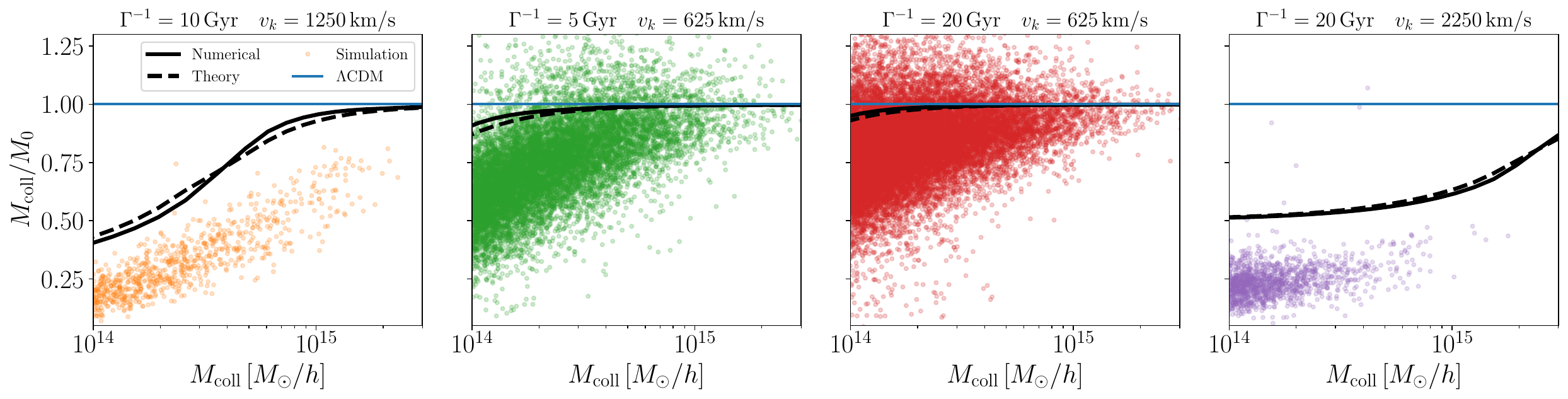}
    \caption{Retained mass fraction $M_{\rm coll}/M_0$ at $z=0$ as a function of the collapsed mass $M_{\rm coll}$ for the four DDM models (one per panel). Solid black lines show the full numerical solution of the collapse system; dashed black lines show the semi-analytical approximation of eqs.~\eqref{eq:mcollM0} and~\eqref{eq:fbound_analytic} evaluated along the EdS cycloid; the blue horizontal line marks the $\Lambda$CDM limit $M_{\rm coll}=M_0$. Points show the per-halo mass ratio $M_{200m}^{\rm DDM}/M_{200m}^{\Lambda{\rm CDM}}$ measured by cross-matching individual haloes between each DDM simulation and the $\Lambda$CDM run, which share identical initial conditions.}
    \label{fig:Mcoll_M0}
\end{figure*}

\subsection{Collapsed mass along the EdS cycloid}
The last ingredient needed for the evaluation of the halo mass function without solving the full system of equations is $M_{\rm coll}$ as required by the Jacobian transformation of eq.~\eqref{eq:hmf_ps_final}. The retained mass fraction is
\begin{equation}\label{eq:mcollM0}
    \frac{M_{\rm coll}}{M_0} = e^{-\Gamma t_{\rm coll}} + \sqrt{1-2\epsilon} 
    \left(1-e^{-\Gamma t_{\rm coll}}\right) \bar{f}_{\rm bound}\,,
\end{equation}
where, using eqs.~\eqref{eq:dotMp} and \eqref{eq:dotMd}, $\bar{f}_{\rm bound}$ reads
\begin{equation}\label{eq:barfbound1}
    \bar f_{\rm bound} = \frac{\Gamma}{1-e^{-\Gamma t_{\rm coll}}} 
    \int_0^{t_{\rm coll}} dt\; f_{\rm bound}(t)\, e^{-\Gamma t}\,.
\end{equation}
This is the fraction of total daughter mass produced during the collapse that remains gravitationally bound to the halo. In principle, evaluating $\bar f_{\rm bound}$ along the EdS cycloid requires $M_{\rm grav} \approx M_0$, i.e.\ that mass loss does not significantly alter the collapse trajectory. In the large-mass limit, this holds to $\mathcal{O}(\epsilon)$ since all daughters are retained and $M_{\rm grav} \approx M_0$ for $\epsilon \ll 1$. In the small-mass limit the trajectory does deviate from EdS since $M_{\rm grav} \approx M_p \ll M_0$, but $f_{\rm bound} \to 0$ in this regime, so the integral in eq.~\eqref{eq:barfbound1} vanishes regardless of the trajectory assumed. The approximation is therefore accurate at both limits, with no assumption required beyond $\epsilon \ll 1$. Using eq.~\eqref{eq:fbound}, equation~\eqref{eq:barfbound1} becomes
\begin{align}\label{eq:fbound_analytic}
    \bar f_{\rm bound} = \frac{\tilde{\Gamma}/\pi}{1-e^{-2\tilde{\Gamma}}} 
    \int_0^{2\pi}d\theta\; (1-\cos\theta)\, 
    f^{\rm EdS}_{\rm bound}(\theta) \\
    \exp\left[-\tilde{\Gamma}\frac{\theta - \sin\theta}{\pi}\right]\nonumber\,,
\end{align}
where $f^{\rm EdS}_{\rm bound}$ is eq.~\eqref{eq:fbound} evaluated on the cycloid using
\begin{align}\label{eqs:beta_xi_eds}
    \omega_{\rm EdS}(\theta) &= 
        \frac{\pi}{t_{\rm ta}}\,(1-\cos\theta)^{-3/2}\,,\nonumber\\
    \xi_{\rm EdS}(\theta) &= 
        \frac{2 v_k t_{\rm ta}}{\pi R_{\rm ta}}\,\sqrt{1-\cos\theta}\,,
        \nonumber\\
    \beta_{\rm EdS}(\theta) &= 
        \frac{|\sin\theta|}{\sqrt{1-\cos\theta}}\,.
\end{align}
Note that eq.~\eqref{eq:fbound_analytic} is correctly normalised: when $f_{\rm bound}^{\rm EdS} = 1$, we have $\bar f_{\rm bound} = 1$ exactly. 
In Fig.~\ref{fig:Mcoll_M0}, we show $M_{\rm coll}/M_0$ as a function of $M_{\rm coll}$ for the four DDM models. The dashed lines (``Theory'') use the semi-analytical expression eq.~\eqref{eq:mcollM0}, while the solid lines (``Numerical'') are obtained by solving the full ODE system. The two agree to better than $7\%$ across the full mass range, with residuals largest for the most extreme models. The colored points show the ratio measured halo-by-halo in the N-body simulations and will be discussed in Sec.~\ref{sec:validation}.

\section{N-body simulation}
\label{sec:simulations}

We compare our theoretical HMF predictions against a suite of N-body simulations using the same implementation as \citet{Bucko:2023eix}. The simulations are run with the \texttt{PKDGRAV3} code \citep{Potter:2016ttn}, a tree-based gravity solver with fast multipole expansion and adaptive time stepping. DDM is implemented by applying stochastic velocity kicks to dark matter particles at each global integration timestep, with kick probability $P = \Gamma \Delta t$ over a timestep $\Delta t$. Since the kick velocities considered here satisfy $v_k \ll c$, the associated change in the homogeneous expansion is subleading, and the standard \texttt{PKDGRAV3} background implementation is retained.

All simulations share the same fiducial cosmology, chosen to match the original \citet{Bucko:2023eix} emulator for consistency: $h = 0.6776$, $\Omega_m = 0.307$, $\Omega_\Lambda = 0.693$, $\sigma_8 = 0.8825$, and $n_s = 0.9665$. Each run uses $512^3$ particles in a comoving box of side length $L = 1000\,h^{-1}{\rm Mpc}$, starting from $z = 49$. We run one $\Lambda$CDM reference simulation and four DDM simulations, covering decay times $\Gamma^{-1} \in \{5, 10, 20\}$~Gyr and velocity kicks $v_k \in \{625, 1250, 2250\}$~km/s. 

Halo catalogues are extracted using AMIGA Halo Finder \citep{Knollmann:2009pb} at two redshifts: $z = 0$ and $z \approx 1.08$. For each halo, the finder provides the mass $M_{200\mathrm{c}}$ and radius $R_{200\mathrm{c}}$, defined as the mass and radius enclosing a mean density equal to $200\,\rho_{\mathrm{c}}(z)$, where $\rho_{\mathrm{c}}$ is the critical density. It additionally fits a Navarro--Frenk--White (NFW) profile \citep{Navarro:1996gj} to each halo,
\begin{equation}
    \rho(r) = \frac{\rho_s}{\left(r/r_s\right)\left(1 + r/r_s\right)^2},
\end{equation}
where $\rho_s$ is a characteristic density and $r_s$ is the scale radius. The concentration parameter $c_{200\mathrm{c}} \equiv R_{200\mathrm{c}}/r_s$ is thus measured individually for each halo from the profile fit. Since the PS formalism is built on a Lagrangian mass $M_0 = (4\pi/3)R_0^3\,\bar{\rho}_m$, the natural final halo-mass definition for comparison with the simulation counterpart is $M_{200\mathrm{m}}$, defined as the mass enclosed within a radius where the mean interior density equals $200\,\bar{\rho}_m(z)$. This identification is motivated by spherical collapse theory \citep{Desjacques:2016bnm}. We therefore convert all halo masses from $M_{200\mathrm{c}}$ to $M_{200\mathrm{m}}$ using the public code \texttt{Colossus} \citep{Diemer:2017bwl}.

\section{Validation}\label{sec:validation}
\begin{figure*}
    \centering
    \includegraphics[width=1\linewidth]{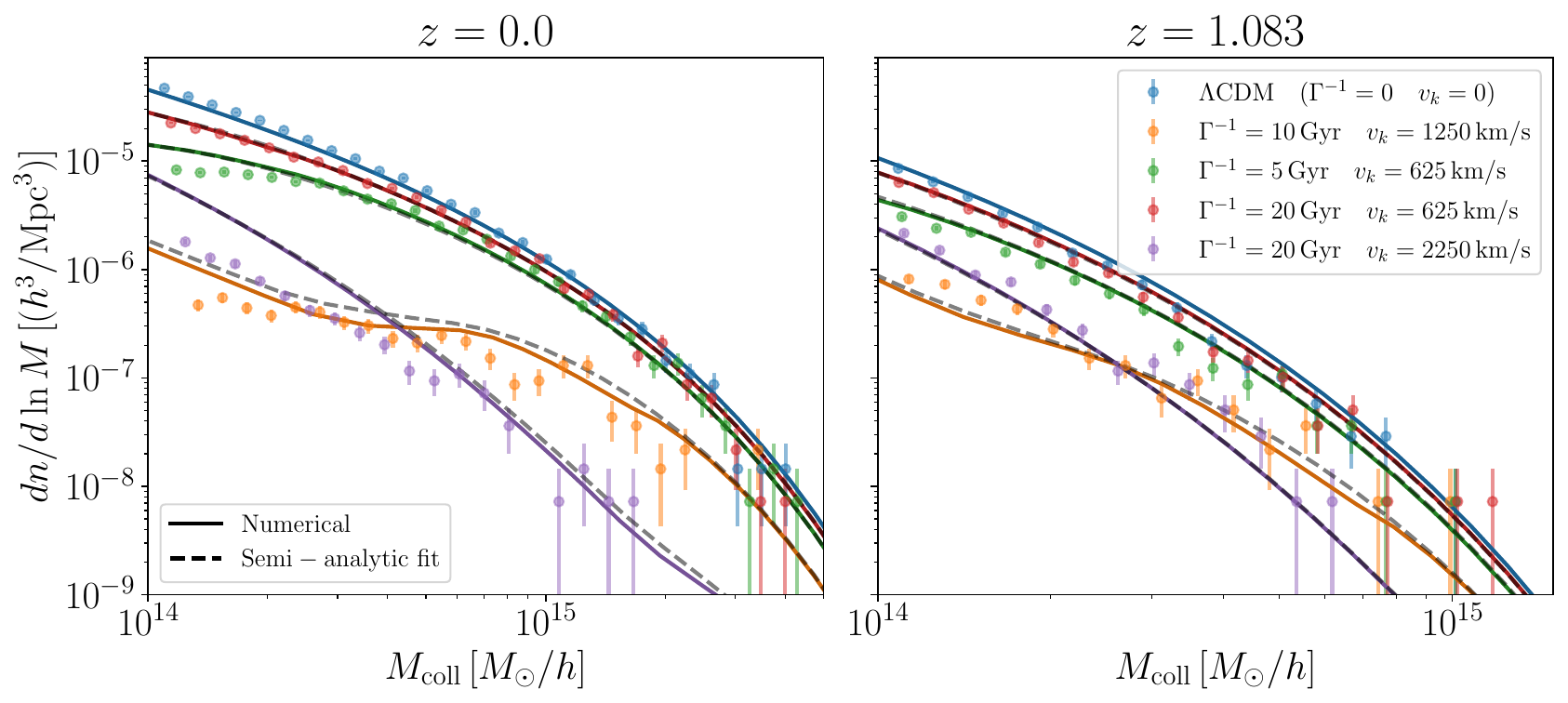}
    \caption{Halo mass function for $\Lambda$CDM and four DDM models at $z=0$ (left) and $z=1.083$ (right). Solid lines show the theoretical predictions from the modified Press--Schechter formalism developed in Sec.~\ref{sec:theory}, while dashed lines show the semi-analytic fit obtained in Sec.~\ref{sec:Critical_density}. Points with error bars show the corresponding N-body simulation measurements. Colors distinguish the five models as indicated in the legend. Note that the $\Lambda$CDM curve is obtained through the same pipeline as the DDM models by setting $\Gamma=0$~Gyr and $v_k=0$~km$/$s.}
    \label{fig:HMF_result}
\end{figure*}

We show in Fig.~\ref{fig:HMF_result} the halo mass function at $z=0$ and $z=1.083$ for $\Lambda$CDM and the four DDM models listed in the legend, comparing our predictions obtained numerically by solving the full system of ODE (solid), and using the semi-analytic fit of sec.~\ref{sec:Critical_density} (dashed), against the N-body simulations (points). We first note that our pipeline correctly recovers the $\Lambda$CDM halo mass function at both redshifts, validating the implementation. We also note that the semi-analytic fit in dashed gray lines reproduces well the numerical HMF. 

Turning to the DDM models, the overall level of suppression relative to $\Lambda$CDM increases with shorter lifetime $\Gamma^{-1}$ and larger kick velocity $v_k$. At $z=1.083$, our predictions remain in good agreement with the simulations across all considered models, spanning from mild deviations from $\Lambda$CDM in red and green (low velocity kicks) to strong suppression in purple and orange (large velocity kicks). 
At $z=0$, our theoretical predictions reproduce the shape and amplitude of the suppression seen in the simulations for the two models with $v_k = 625\,\mathrm{km/s}$. For the models with large velocity kicks ($v_k = 1250$ and $2250\,\mathrm{km/s}$), our model overestimates the abundance of halos, while still capturing the qualitative shape of the HMF. We argue below that this discrepancy originates in the identification of the collapsed mass $M_{\rm coll}$ with the halo-finder mass $M_{200m}$, rather than in the collapse dynamics itself.

The origin of the residual discrepancy for the large-kick models can be traced with the halo-by-halo comparison of Fig.~\ref{fig:Mcoll_M0}. Since the DDM and $\Lambda$CDM simulations share identical initial conditions, each massive halo can be cross-matched between the runs, and the measured ratio $M_{200m}^{\rm DDM}/M_{200m}^{\Lambda{\rm CDM}}$ provides a measurement of the mass mapping that our framework models through $M_{\rm coll}(M_0)$. For all four models, the predicted $M_{\rm coll}/M_0$ traces the upper envelope of the matched ratios, as expected by construction: $M_{\rm coll}$ counts all bound material, while the halo finder only registers particles instantaneously located within $R_{200m}$. For the two models with $v_k = 625\,\mathrm{km/s}$, the matched ratios lie close below the prediction, and consistently the HMF is well reproduced at both redshifts. For the two large-kick models, however, the measured ratios fall systematically below the prediction, by up to a factor of two at $M \sim 10^{14}\,M_\odot/h$, and the HMF is overestimated.

This behaviour reflects a genuine ambiguity in the definition of the halo mass. Our $M_{\rm coll}$ counts the surviving parent mass plus all gravitationally bound daughter particles, and is evaluated in the formal collapse limit $R \to 0$, where the potential well is arbitrarily deep. The halo finder, in contrast, only counts particles instantaneously located within the finite radius $R_{200m}$ at the snapshot time. The two definitions coincide when daughter orbits are small compared to the halo. For the large-kick models, instead, the kick velocity is comparable to or exceeds the halo circular velocity. Retained daughters then belong almost entirely to population 2 of Sec.~\ref{sec:theory}: bound, but on orbits extending well beyond the halo boundary. Such particles spend a large fraction of each orbital period outside $R_{200m}$ and are therefore only partially registered by the halo finder, while being fully counted in $M_{\rm coll}$.

This effect could be incorporated within the present framework by weighting population-2 daughters by the fraction of their orbital period spent inside $R$, which is computable in the harmonic potential of Sec.~\ref{sec:theory}, thereby defining a finder-consistent ``observable mass''. We leave this refinement to future work, noting that it significantly affects only extreme models in which the kick velocity exceeds the circular velocity of all haloes in the sample and that even there, the prediction correctly captures the qualitative shape of the suppression.

\section{Conclusions}
\label{sec:conclusions}
In this work, we have developed a self-consistent semi-analytical framework for the halo mass function in decaying dark matter cosmologies, where the parent particle decays into one massive daughter carrying a velocity kick $v_k$ and one massless dark radiation component. The simpler scenario in which the parent decays entirely into dark radiation is naturally encompassed within the same framework.
Our approach builds on the Press--Schechter formalism, improved by a spherical collapse model that explicitly tracks the mass loss induced by the decay of dark matter particles. We showed that injecting the DDM linear power spectrum into the standard PS framework fails to reproduce the simulated HMF regardless of the window function adopted. Hence, keeping the $\Lambda$CDM power spectrum to compute the variance $\sigma(M)$, we encoded all DDM physics through a modified, mass-dependent critical density $\delta_c(M)$ obtained via spherical collapse.

The key physical ingredient is the partition of daughter particles into three populations: fully bound and fully-interior to the halo, bound but with orbits crossing the halo boundary, and unbound. Unbound daughter particles are removed from the gravitating mass instantaneously after creation, an approximation we refer to as instantaneous escape. The kinematics of the velocity kick yield closed-form analytical expressions for the interior fraction $f_{\rm in}$ and the bound fraction $f_{\rm bound}$. For the intermediate population, i.e. bound daughter particles whose orbits extend beyond the halo boundary, we introduce an effective gravitating mass $M_{\rm grav}$ that smoothly interpolates between parent-dominated and daughter-dominated regimes, capturing the progressive ``puffing'' of the halo as bound daughter particles build up beyond $R$.

To test our theoretical prediction, we ran a suite of four N-body simulations using the code developed in \citet{Bucko:2023eix}: two models close to $\Lambda$CDM with small velocity kicks, $v_k = 625$~km$/$s, and lifetimes $\Gamma^{-1} = 5$~Gyr and $\Gamma^{-1} = 20$~Gyr; and two models with large velocity kicks yielding stronger suppression, $v_k = 1250$~km$/$s with $\Gamma^{-1} = 10$~Gyr and $v_k = 2250$~km$/$s with $\Gamma^{-1} = 20$~Gyr. At $z \approx 1.08$, we find good agreement across all four models, and at $z = 0$ for the two models with $v_k = 625$~km$/$s. For the two large-kick models at $z = 0$, the predicted abundance is overestimated at low masses. By cross-matching individual haloes between the DDM and $\Lambda$CDM simulations, which share identical initial conditions, we traced this discrepancy to the identification of the collapsed mass with the halo-finder mass: $M_{\rm coll}$ counts all gravitationally bound material, whereas $M_{200m}$ only registers particles instantaneously located within $R_{200m}$.
The two definitions diverge precisely when the kick velocity exceeds the halo circular velocity in which case retained daughters' orbits extend beyond the halo boundary. An ``observable mass'' can be defined within the same framework by weighting daughters by the fraction of their orbital period spent inside $R$. We note that it significantly affects only an extreme corner of the DDM parameter space already in strong tension with existing constraints \citet{Montandon:2025xpd, Bucko:2023eix}.

Computing $\delta_c$ for an arbitrary DDM model requires solving the full system of equations: gravitational dynamics~\eqref{eq:EOM}, dark matter decay~\eqref{eq:dotMp} and~\eqref{eq:dotMd}, and halo puffing~\eqref{eq:Mgrav}. While accurate, this numerical approach may be expensive for cosmological parameter inference. We therefore provide a systematic study of $\delta_c(M)$ and derive analytical expressions in the large- and small-mass regimes, together with a fitting formula for the transition that depends on a single free parameter, the characteristic mass scale $M_1$.
The critical density $\delta_c(M)$ exhibits a mass-dependent transition between two analytically tractable plateaus, visible for all models in Fig.~\ref{fig:delta_c}. In the large-mass limit, where all daughter particles are retained, we derive a closed-form perturbative expression, Eq.~\eqref{eq:deltac_large}, accurate to better than $1\%$ over the full range of models considered. In the small-mass limit, the collapse dynamics becomes equivalent to a pure dark radiation decay and $\delta_c$ depends on $\tilde{\Gamma}$ alone, well described by the fitting formula of Eq.~\eqref{eq:deltac_plateau}, accurate to $1.5\%$. The transition between these two regimes is captured by the fitting function of Eq.~\eqref{eq:fitting_function}, whose shape parameters are universal across all DDM models considered, and whose single free mass scale $M_1$ follows the analytical scaling $M_1 \propto v_k^3\,\tilde{\Gamma}^{-1/2}t_{\rm ta}$. Finally, we also provide a semi-analytical formula to compute the collapsed mass $M_{\rm coll}$ as function of the initial Lagrangian mass $M_0$ in \eqref{eq:mcollM0}.

Beyond its role as a physical model, our framework is built for practical use in cosmological inference. The analytic large- and small-mass plateaus and the single-parameter transition fit reduce the full collapse calculation to a closed-form prescription for $\delta_c(M)$ that can be evaluated at negligible cost inside an MCMC pipeline, making a joint analysis of the DDM lifetime and kick velocity with the halo abundance tractable. This is timely given the cluster mass function is now being measured with percent-level precision.

The most immediate application is to cluster number counts. X-ray samples from eROSITA (eRASS), thermal Sunyaev--Zel'dovich catalogues from SPT, ACT and the Simons Observatory --- and, in the coming decade, CMB-S4 --- and optically or weak-lensing selected clusters from DES, \textit{Euclid} and the Vera C.\ Rubin Observatory (LSST) collectively span the mass and redshift range where our predicted suppression is largest. With a closed-form $\delta_c(M)$ in hand, we are now equipped to compute the DDM cluster mass function across mass and redshift and to derive constraints on the decay lifetime and kick velocity from these datasets, offering an independent cross-check on the DDM interpretation of the $S_8$ tension \citep{Abellan:2020pmw,Abellan:2021bpx,Bucko:2023eix,Montandon:2025xpd}. A further feature of the DDM signal is that it grows towards low redshift and high mass, so that the redshift evolution of the cluster mass function carries additional information beyond its amplitude at a single epoch. Confronting our HMF with these datasets will be the subject of future work.

\section*{Acknowledgments}
The authors thank gratefully Elsa Teixeira and Julien Lavalle for useful discussions. TM and VP are supported by funding from the European Research Council (ERC) under the European Union’s HORIZON-ERC-2022 (grant agreement no. 101076865). TM and VP acknowledge the European Union's Horizon Europe research and innovation programme under the Marie Sk\l odowska-Curie Staff Exchange grant agreement no.\ 101086085 -- ASYMMETRY.

\appendix 
\section{Einstein-de Sitter cycloid}\label{app:EdS}
We give here a brief review of the standard spherical collapse model in an Einstein-de Sitter universe, following Ref.~\citep{Desjacques:2016bnm}. In this case, the halo mass is constant over time and the shell radius obeys
\begin{equation}\label{eq:EdScycloid}
    \ddot R = -\frac{G M_0}{R^2}\,.
\end{equation}
It is common to define the turnaround coefficients $t_{\rm ta}$ and $R_{\rm ta}$ through Kepler's relation
\begin{equation}\label{eq:app:Kepler}
    G M_0  = \frac{\pi^2}{8} \frac{R_{\rm ta}^3}{t_{\rm ta}^2}\,.   
\end{equation}
Using the dimensionless variables $\tilde R = R / R_{\rm ta}$ and $\tilde t = t / t_{\rm ta}$, Eq.~\eqref{eq:EdScycloid} becomes Eq.~\eqref{eq:EOM} with $M_{\rm grav}=M_0$. The solution can then be written in parametric form as
\begin{equation}\label{eq:app:general_sol}
    \tilde R_{\rm EdS}(\theta) = \frac{1}{2} (1-\cos \theta)\,,\qquad 
    \tilde t_{\rm EdS}(\theta) = \frac{1}{\pi} (\theta-\sin \theta)\,.
\end{equation}

\section{Large-mass limit derivation}\label{app:largemasslimit}
We now solve the large-mass limit equation in DDM
\begin{align}\label{eq:app:largemass_ODE2}
    \tilde R_1'' -\frac{2\pi^2}{8 \tilde R^3_{\rm EdS}} \tilde R_1
    &= S(\tilde t)\,,\\
    S(\tilde t)
    &\equiv -\frac{\pi^2}{8 \tilde R_{\rm EdS}^2}
    \left(e^{-\tilde \Gamma \tilde t} - 1\right)\,.
\end{align}
This differential equation admits two homogeneous solutions. Differentiating the zeroth-order EdS solution gives one of them,
\begin{equation}
    \tilde R_1^{(1)} = \tilde R_{\rm EdS}' = \frac{\pi \sin \theta}{2(1-\cos \theta)}\,,
\end{equation}
where the prime denotes a derivative with respect to $\tilde t$. Using reduction of order, we write $\tilde R^{(2)}_1 = \tilde R^{(1)}_1 f(\tilde t)$. Injecting this ansatz in Eq.~\eqref{eq:app:largemass_ODE2} and using the homogeneous equation obeyed by $\tilde R_1^{(1)}$, we find
\begin{align}
    \tilde R_1^{(2)} &= \tilde R_1^{(1)} \int^\theta_0  \frac{d\tilde t}{d\theta'} \frac{d\theta'}{\tilde R^{(1)2}_{1}(\theta')}\\
    &=\frac{4}{\pi^3} \tilde R_1^{(1)} \int_0^{\theta}d\theta'\frac{(1-\cos\theta')^3}{\sin^2\theta'}\\
    &=\frac{4}{\pi^3}\tilde R_1^{(1)} \left[\sin\theta-3 \theta  + 4 \tan \frac{\theta}{2}\right]\\
    &=\frac{4}{\pi^3}\tilde R_1^{(1)} I(\theta)\,.
\end{align}
The particular solution obtained by variation of parameters is
\begin{align}
    \tilde R^{\rm part}_1(\theta) =&
    \frac{\tilde R_1^{(2)}(\theta)}{W}
    \int_0^{\theta}d\theta'\,
    \tilde R_1^{(1)}(\theta') S(\theta') \frac{d\tilde t}{d \theta'} \nonumber\\
    &-
    \frac{\tilde R_1^{(1)}(\theta)}{W}
    \int_0^{\theta}d\theta'\,
    \tilde R_1^{(2)}(\theta') S(\theta') \frac{d\tilde t}{d \theta'}\,,
\end{align}
where $W = \tilde R_1^{(1)} \tilde R_1^{(2)'} - \tilde R_1^{(2)} \tilde R_1^{(1)'}$ is the Wronskian. With the normalization chosen above, $W=1$. The particular solution can be simplified to
\begin{align}\label{app:part}
    \tilde R^{\rm part}_1(\theta)
    =& \frac{1}{\pi}\tilde R_1^{(1)}(\theta)
    \int_0^\theta d\theta'\,
    \frac{\sin\theta'\left(1 - e^{-\tilde\Gamma \tilde t(\theta')}\right)}
    {(1-\cos\theta')^2}
    \left[I(\theta) - I(\theta')\right]\,.
\end{align}
The particular solution \eqref{app:part} dominates over the homogeneous solutions, whose amplitudes are set by initial conditions at $t_0 \ll t_{\rm coll}$ before any significant decay has occurred, and are therefore suppressed by $(t_0/t_{\rm coll})^{2/3} \ll 1$.

We now determine the collapse-time shift by asymptotic matching near the EdS singularity. Writing the time to EdS collapse as $\tau_{\rm EdS} = 2-\tilde t$, the expansion of the particular solution near $\theta=2\pi$ gives
\begin{align}\label{app:nongeneric_form_app}
    \tilde R(\tau_{\rm EdS}) &= \tilde R_{\rm EdS}(\tau_{\rm EdS}) + \epsilon \tilde R^{\rm part}_1(\tau_{\rm EdS}) \nonumber\\
    &\approx \frac{(6\pi)^{2/3}}{4}\tau_{\rm EdS}^{2/3}
    -\frac{\epsilon J(\tilde \Gamma)}{(6\pi)^{1/3}}\tau_{\rm EdS}^{-1/3}\,,
\end{align}
where $J(\tilde \Gamma)$ is defined in Eq.~\eqref{eq:J}. 

We match this expression to the universal spherical-collapse form near collapse. At this order, the equation reduces to $\tilde R'' \approx -\pi^2/(8\tilde R^2)$ up to corrections that only affect the normalization at higher order, so the local solution is $\tilde R \propto (\tilde t_{\rm coll}-\tilde t)^{2/3}$. Defining $\tilde t_{\rm coll} = 2 + \delta\tilde t$, we have
\begin{equation}
    \tilde R = \frac{(6\pi)^{2/3}}{4} \left(\tau_{\rm EdS} + \delta\tilde t\right)^{2/3}\,,
\end{equation}
which expands to
\begin{equation}\label{app:generic_form}
    \tilde R = \frac{(6\pi)^{2/3}}{4} \tau_{\rm EdS}^{2/3}
    + \frac{(6\pi)^{2/3}}{6} \delta\tilde t \tau_{\rm EdS}^{-1/3}\,.
\end{equation}
Matching the coefficients of $\tau_{\rm EdS}^{-1/3}$ in Eqs.~\eqref{app:nongeneric_form_app} and \eqref{app:generic_form} gives Eq.~\eqref{eq:deltat}. The linear extrapolation to the actual collapse time then leads to Eq.~\eqref{eq:deltac_large}.

\bibliography{main_inspirehep}

\end{document}